%% file: main.tex
\documentclass[lettersize,journal]{IEEEtran}
\usepackage[T1]{fontenc}
\usepackage[latin9]{inputenc}
\usepackage{array}
\usepackage{ifsym}
\usepackage{multirow}
\usepackage{amstext}
\usepackage{amsthm}
\usepackage{stackrel}
\usepackage{graphicx}
\usepackage[unicode=true,
 bookmarks=true,bookmarksnumbered=true,bookmarksopen=true,bookmarksopenlevel=1,
 breaklinks=false,pdfborder={0 0 0},pdfborderstyle={},backref=false,colorlinks=false]
 {hyperref}
\hypersetup{pdftitle={Your Title},
 pdfauthor={Your Name},
 pdfpagelayout=OneColumn, pdfnewwindow=true, pdfstartview=XYZ, plainpages=false}
\usepackage{breakurl}
\usepackage{color}
\usepackage{enumitem}
\renewcommand\color[2]{#2}

\makeatletter

\DeclareRobustCommand*{\lyxarrow}{%
\@ifstar
{\leavevmode\,$\triangleleft$\,\allowbreak}
{\leavevmode\,$\triangleright$\,\allowbreak}}
\newcommand{\binom}[2]{{#1 \choose #2}}

\providecommand{\tabularnewline}{\\}

\theoremstyle{plain}
\newtheorem{thm}{\protect\theoremname}
\theoremstyle{plain}

\theoremstyle{definition}
\newtheorem{definition}{Definition}

\usepackage[caption=false,font=footnotesize]{subfig}

\@ifundefined{showcaptionsetup}{}{%
 \PassOptionsToPackage{caption=false}{subfig}}
\usepackage{subfig}
\makeatother

\providecommand{\lemmaname}{Lemma}
\providecommand{\theoremname}{Theorem}

\begin{document}

\title{\vspace{-0pt} SP-Chain: Boosting Intra-Shard and Cross-Shard Security and Performance in Blockchain Sharding}


\author{Mingzhe~Li,~\IEEEmembership{Member,~IEEE,}
        You~Lin,
        Wei~Wang,~\IEEEmembership{Member,~IEEE,}
        and~Jin~Zhang,~\IEEEmembership{Member,~IEEE}
\IEEEcompsocitemizethanks{
\IEEEcompsocthanksitem This work was supported by the National Natural Science
Foundation of China (NSFC) under Grant T2495254.
\IEEEcompsocthanksitem M. Li is with the Department of Computer Science and Engineering, Southern University of Science and Technology, Shenzhen, China, and also with the Department
of Computer Science and Engineering, Hong Kong University of Science and Technology, Hong Kong
(email: mlibn@connect.ust.hk).
\IEEEcompsocthanksitem Y. Lin is with the Department of Computer Science and Engineering, Southern University of Science and Technology, Shenzhen, China (email: liny2021@mail.sustech.edu.cn).
\IEEEcompsocthanksitem W. Wang is with the Department
of Computer Science and Engineering, Hong Kong University of Science and Technology, Hong Kong (email: weiwa@cse.ust.hk).
\IEEEcompsocthanksitem J. Zhang is with the Department of Computer Science and Engineering
and the Research Institute of Trustworthy Autonomous Systems, Southern
University of Science and Technology, Shenzhen 518055, China, and also
with Peng Cheng Laboratory, Shenzhen 518055, China (email: zhangj4@sustech.edu.cn).
\IEEEcompsocthanksitem J. Zhang and W. Wang are the corresponding authors.
}
}

\markboth{IEEE Internet of Things Journal, VOL. XX, NO. XX, XX 2024}%
{Li \MakeLowercase{\textit{et al.}}: SP-Chain: Boosting Intra-Shard and Cross-Shard Security and Performance in Blockchain Sharding}

\IEEEtitleabstractindextext{%
\begin{abstract}
A promising way to overcome the scalability limitations of the current blockchain is to use sharding, which is to split the transaction processing among multiple, smaller groups of nodes.
A {\color{red}well-performing} blockchain sharding system requires both \emph{high performance and high security in both intra- and cross-shard perspectives.}
However, existing protocols either have {\color{red}issues in protecting security} or trade off great performance for security.
In this paper, we propose SP-Chain, a blockchain sharding system with enhanced \underline{S}ecurity and \underline{P}erformance for both intra- and cross-shard perspectives.
For the intra-shard aspect, we design a pipelined two-phase concurrent voting scheme to provide high system throughput and low transaction confirmation latency.  
Moreover, we propose an efficient unbiased leader rotation scheme to ensure high performance under malicious behavior. 
For the cross-shard aspect, a proof-assisted efficient cross-shard transaction processing mechanism is proposed to guard cross-shard transactions with low overhead.
We implement SP-Chain based on Harmony, and evaluate its performance via large-scale deployment.
Extensive evaluations suggest that SP-Chain can process more than 10,000 tx/sec under malicious behaviors with a confirmation latency of 7.6s in a network of 4,000 nodes.

\end{abstract}

\begin{IEEEkeywords}
Blockchain, blockchain sharding, cross-shard transaction processing, intra-shard transaction processing
\end{IEEEkeywords}}

\maketitle

\IEEEdisplaynontitleabstractindextext

\IEEEpeerreviewmaketitle

\input{IntroV2}

\input{background}

\input{model}

\input{design}

\input{analysis}

\input{evaluation}

\input{Discussion}

\section{Conclusions}

We present SP-Chain, a sharding-based blockchain system with scalability, high throughput, low latency and reliable security.
We exploit blockchain sharding systems' features and design an intra-shard consensus protocol called concurrent voting for the sharding system. 
This protocol can significantly improve system performance. 
Based on this protocol, we propose an unbiased leader rotation scheme. 
It can help the system maintain high efficiency in the presence of malicious behaviors. 
An efficient and verifiable cross-shard transaction processing mechanism ensures the security of cross-shard transactions.
We have implemented a prototype of SP-Chain. 
Our empirical evaluation demonstrates that SP-Chain scales smoothly to network sizes of up to 4,000 nodes showing better performance than previous works.

\bibliographystyle{abbrv}
\bibliography{reference}

\begin{IEEEbiography}[{\includegraphics[width=1in,height=1.25in,clip,keepaspectratio]{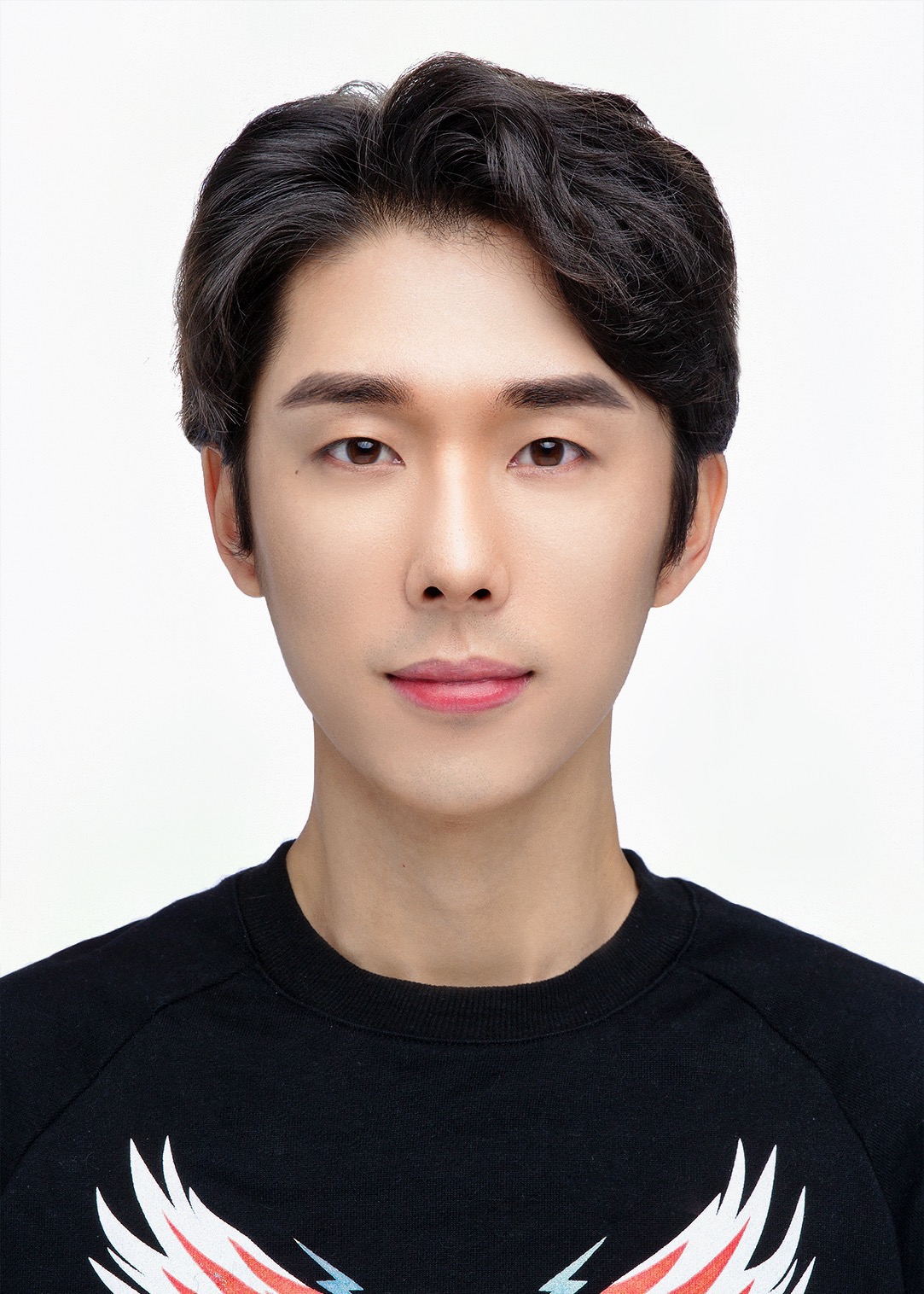}}]{Mingzhe Li}
is currently a Scientist with the Institute of High Performance Computing (IHPC), A*STAR, Singapore.
He received his Ph.D. degree from the Department of Computer Science and Engineering, Hong Kong University of Science and Technology in 2022.
Prior to that, he received his B.E. degree from Southern University of Science and Technology.
His research interests are mainly in blockchain sharding, consensus protocol, blockchain application, network economics, and crowdsourcing.
\end{IEEEbiography}
\vspace{-30pt}
\begin{IEEEbiography}[{\includegraphics[width=1in,height=1.25in,clip,keepaspectratio]{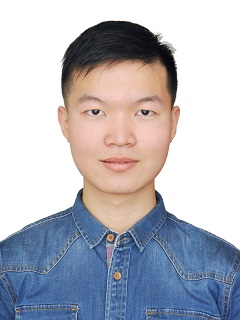}}]{You Lin}
is currently a master candidate with Department of Computer Science and Engineering, Southern University of Science and Technology. 
He received his B.E. degree in computer science and technology from Southern University of Science and Technology in 2021. 
His research interests are mainly in blockchain, network economics, and consensus protocols.
\end{IEEEbiography}
\vspace{-30pt}
\begin{IEEEbiography}
[{\includegraphics[width=1in,height=1.25in,clip,keepaspectratio]{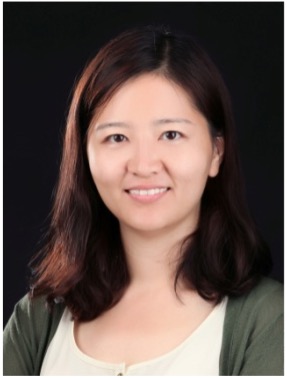}}]{Jin Zhang} 
is currently an associate professor with Department of Computer Science and Engineering, Southern University of Science and Technology. 
She received her B.E. and M.E. degrees in electronic engineering from Tsinghua University in 2004 and 2006, respectively, and received her Ph.D. degree in computer science from Hong Kong University of Science and Technology in 2009. 
Her research interests are mainly in mobile healthcare and wearable computing, wireless communication and networks, network economics, cognitive radio networks and dynamic spectrum management. 
\end{IEEEbiography}
\vspace{-30pt}
\begin{IEEEbiography}
[{\includegraphics[width=1in,height=1.25in,clip,keepaspectratio]{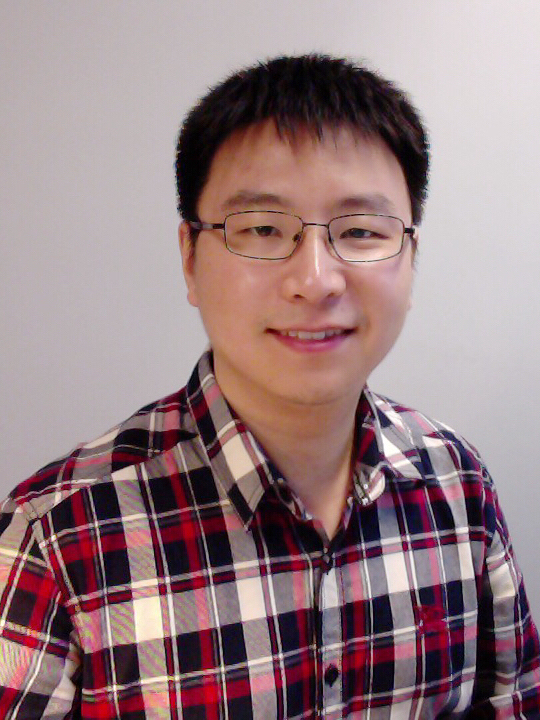}}]{Wei Wang}
is currently an associate professor with Department of Computer Science and Engineering, Hong Kong University of Science and Technology.
He received the Ph.D. degree from the Department of Electrical and Computer Engineering, University of Toronto in 2015.
Prior to that, he obtained the B.Eng. and M.Sc. degrees from Shanghai Jiao Tong University. 
He is also affiliated with HKUST Big Data Institute.
His research interests cover the broad area of networking and distributed systems, with a special focus on big data and machine learning systems, cloud computing, and computer networks. 
\end{IEEEbiography}

\end{document}

%% file: IntroV2.tex
\section{Introduction}

\IEEEPARstart{S}{ince}
the advent of Bitcoin \cite{nakamoto2008bitcoin}, blockchain systems have continued to have a significant impact on society. 
However, the low system throughput and high transaction confirmation delays of such systems greatly hinder their usability across various infrastructures and applications. 
Consequently, sharding \cite{corbett2013spanner, luu2016secure, zamani2018rapidchain, al2024sok}, a promising blockchain scaling solution, has been proposed. 
Herein, the entire blockchain state is divided into multiple non-overlapping shards, each maintained by a group of nodes.

\emph{Performance and security} are critical areas of concern in blockchain sharding systems \cite{luu2016secure, kokoris2018omniledger}. 
To improve sharding performance and security, considerations need to be made from both \emph{intra-shard and cross-shard} perspectives. 
In blockchain sharding, each shard must not only handle transactions within the shard but also transmit a large number of cross-shard transactions to other shards \cite{zamani2018rapidchain, wang2019monoxide, hong2021pyramid, li2023lb}. 
However, as will be discussed below, existing protocols either have issues ensuring security within and between shards or sacrifice significant performance for security.

\vspace{3pt}
\noindent
\textbf{Intra-shard security and performance:}
Within each shard, a specific leader is typically selected to propose blocks. 
\emph{Malicious leaders or attacks on leaders can severely affect intra-shard security and performance. }
Specifically, most existing sharding protocols \cite{luu2016secure, kokoris2018omniledger, zhang2020cycledger, hong2021pyramid, li2022jenga, li2023cochain} require each shard's leader to produce blocks within a certain time frame and to know the next leader in advance. 
This allows attackers to easily launch \emph{targeted attacks on the leaders} \cite{li2020gosig}, weakening system security (for example, colluding with the leader for Byzantine behavior or launching {\color{red}DDoS (Distributed Denial-of-Service)} attacks against the leader). 
Moreover, when a leader is found to be malicious or under attack, a complex \emph{view change} process is required to elect a new leader \cite{castro2002practical}, thereby reducing system performance.

Another issue is that many existing sharding protocols adopt generic Byzantine Fault Tolerance (BFT) consensus protocols for intra-shard consensus, requiring \emph{extensive communication among shard members.} 
Specifically, such consensus protocols {\color{red}(e.g., PBFT, Practical Byzantine Fault Tolerance)} \cite{castro2002practical} used in partial-synchronous networks typically require multi-stage communication (e.g., pre-preparation, preparation, commit) to ensure security in poor network connectivity and high latency situations. 
However, these protocols are inefficient when applied to blockchain sharding. 
In sharding systems, each shard generally has high network connectivity and a low and fixed message propagation cap \cite{zamani2018rapidchain, kokoris2018omniledger, huang2020repchain, zhang2020cycledger, li2020polyshard}. 
Under such conditions, a more efficient consensus protocol with less communication overhead can be proposed for sharding systems.

\vspace{3pt}
\noindent
\textbf{Cross-shard security and performance:}
Many cross-shard transactions occur in blockchain sharding systems. 
\emph{Malicious leaders may send incorrect cross-shard transactions to other shards \cite{zhang2020cycledger}. }
Because each shard maintains information isolation, the shard receiving the transaction cannot verify the validity of the cross-shard transaction, endangering system security. 
However, some existing sharding protocols ignore how to ensure the security of cross-shard transactions \cite{zamani2018rapidchain, kokoris2018omniledger, hong2021pyramid}. 
Other solutions \cite{wang2019monoxide, zhang2020cycledger, li2023cochain} require each cross-shard transaction to be accompanied by a large amount of additional proof information to ensure security, bringing excessive overhead to the system.

\vspace{3pt}
To simultaneously improve intra-shard and cross-shard performance and security, the following \textbf{\emph{challenges}} need to be addressed. 
\emph{First}, how to design a secure and efficient leader election protocol that can both resist attacks on leaders and efficiently elect new leaders. 
\emph{Second}, how to design an efficient intra-shard consensus protocol that takes advantage of the characteristics of sharding systems. 
\emph{Third}, how to design efficient and secure cross-shard transaction processing mechanisms. 
To address these challenges, we propose the following design points:

\vspace{3pt}
\noindent
\textbf{Random Leader Rotation. }
To address the first challenge, the leader of each shard is frequently, randomly, and automatically changed. 
Specifically, the leaders are rotated on a per-block cycle.
To prevent attackers from knowing subsequent leaders in advance, leaders are selected based on distributed randomness only before each block is proposed. 
To ensure performance and security during the distributed randomness generation process, we propose a \emph{chain-based randomness generation mechanism}. 
Herein, nodes within each shard use the signature information of previously confirmed blocks to efficiently generate unbiased distributed randomness. 
In addition, the automatic leader rotation mechanism eliminates the originally complex view change process, further improving the efficiency of leader election.

\vspace{3pt}
\noindent
\textbf{Two-Phase Concurrent Voting. }
To address the second challenge, we propose an efficient consensus mechanism suitable for sharding systems with low communication overhead. 
By utilizing the advantages of a small number of nodes, good network connectivity, and a high synchronization rate within each shard, we propose a \emph{synchronous} consensus protocol that compresses the previous preparation and commit stages into one without violating security. 
Therefore, we reduce the 3 rounds of communication required by traditional consensus protocols to 2, thus accelerating the speed of intra-shard consensus.

More importantly, to further enhance consensus efficiency, we {\color{red}\emph{make both the computation and communication steps run in parallel}, whereas in traditional consensus protocols these steps are processed serially}.
In traditional protocols, the leader first packages transactions into a block. 
Then, the leader broadcasts the block. 
Nodes receiving the block will vote and reach a consensus. 
After reaching a consensus, nodes commit the block into the blockchain. 
Our concurrent voting protocol parallelizes these computation and communication steps. 
Specifically, when a shard's leader produces a new block, the members of that shard vote on and reach a consensus on the previous block. 
Also, while the leader broadcasts the new block, each node commits the previous block into the blockchain. 
This proposed scheme provides fast block generation and confirmation rates, bringing high throughput and low latency to the system.

\vspace{3pt}
\noindent
\textbf{Proof-Assisted Efficient Cross-Shard Transaction Processing.}
To address the third challenge, we require cross-shard transactions to \emph{carry batched and pruned proofs for forwarding}. 
Specifically, to allow shards to safely verify the received cross-shard transactions, we require cross-shard transactions to carry proof \cite{liu2021merkle}, demonstrating their validity in the sending shard. 
To reduce the additional overhead that proof brings to network transmission, we propose a batch proof mechanism. 
In this scheme, the transactions sent to the same shard are accompanied by one pruned proof (instead of each transaction requiring separate proof). 
This proof can help nodes batch verify the validity of all transactions sent to the same shard.

{\color{red}
This paper mainly makes the following contributions:

\begin{itemize}[left=0pt]
    \item We propose SP-Chain, a novel blockchain sharding system that significantly improves both intra-shard and cross-shard security and performance.
    \item We design a \emph{pipelined two-phase concurrent voting} protocol for intra-shard consensus, which greatly increases throughput and lowers latency compared to traditional sequential BFT consensus.
    \item We introduce an \emph{unbiased leader rotation} mechanism based on signature-derived randomness that ensures no predictable leader pattern, enhancing security against leader targeting attacks while maintaining high efficiency.
    \item We develop an \emph{efficient proof-assisted cross-shard transaction processing} scheme that securely handles transactions across shards with minor overhead. 
    \item We implement a prototype of SP-Chain based on Harmony\footnote{A well-known public blockchain sharding project that was once ranked in the top 50 in the cryptocurrency space in terms of market capitalization} \cite{Harmony} and conduct extensive large-scale experiments (up to 4,000 nodes). Our evaluation demonstrates that SP-Chain achieves over 10,000 transactions per second (TPS) with low confirmation latency, outperforming state-of-the-art sharding systems.
\end{itemize}
}



%% file: background.tex
\section{Background and Related Work}

\subsection{Blockchain Sharding}

In traditional blockchain protocols \cite{nakamoto2008bitcoin, wood2014ethereum}, all network nodes have to agree on all the transactions. 
This scheme leads to very low throughput and high latency for transactions to be packed into blocks and confirmed.
An alternative way is to partition nodes into disjoint shards and let each shard maintain the states of a subgroup of users \cite{luu2016secure}.
Under this method, the throughput increases proportionally to the number of committees. 
The transaction confirmation latency is also reduced since one committee has fewer nodes.
This technique is known as sharding and is considered an excellent way to help blockchain scale well. 

A {\color{red}well-performing} blockchain sharding system needs to ensure good security as well as high performance. 
Unlike traditional blockchain systems, in blockchain sharding systems, there are a large number of cross-shard transactions that are transmitted among shards \cite{wang2019monoxide, kokoris2018omniledger, zamani2018rapidchain, nguyen2019optchain, huang2022brokerchain, li2023lb, nguyen2022denial}. 
Nodes need to not only process transactions within a shard, but also handle transactions across shards.
Therefore, it is {\color{red}necessary} to consider both intra- and cross-shard aspects when analyzing the security and performance of blockchain sharding.
However, previous works have problems with both intra- and cross-shard security and performance.

\vspace{3pt}
\noindent
\textbf{Issues from Cross-Shard Aspect.}
Cross-shard transaction processing is a significant part of the blockchain sharding system. 
However, existing cross-shard transaction processing schemes still have drawbacks on protecting security, or they increase large amount of communication overhead when securing transactions.
OmniLedger \cite{kokoris2018omniledger} requires an honest client to participate during the cross-shard transaction process. 
A malicious client can lock the cross-shard transactions and obstruct their executions. 
Such behavior brings troubles for both system security and efficiency. 
Some works \cite{zamani2018rapidchain, huang2020repchain, hong2021pyramid, li2022jenga} does not consider the leader to be evil when dealing with cross-shard transactions, making the cross-shard transaction execution less secure. 
Other works \cite{wang2019monoxide, zhang2020cycledger} proposes to attach proofs to each cross-shard transaction to protect security. 
However, this mechanism increases the lots of communication overhead, reducing system efficiency. 
{\color{red}
BrokerChain \cite{huang2024brokerchain} introduces special "broker" accounts to reduce cross-shard transfers in account-based sharding, helping balance the workload across shards. However, its intermediary accounts add protocol complexity.
Authors in \cite{liu2023flexible} propose a flexible sharding protocol that integrates a cross-shard Byzantine fault tolerance (CSBFT) consensus, allowing multiple shards to coordinate on cross-shard transactions in a single consensus round and thereby cutting down confirmation delay. The CSBFT, however, incurs extra cross-shard coordination overhead for each transaction.
LightCross \cite{qi2024lightcross} focuses on minimizing cross-shard communication rounds by using trusted hardware (TEE). Nonetheless, LightCross's reliance on TEE devices introduces additional trust and hardware assumptions.
ShardCon \cite{zhang2024efficient} embeds cross-shard transaction handling into a unified Byzantine consensus spanning the involved shards. 
However, coordinating a BFT consensus among multiple shards for each cross-shard transaction incurs high communication and computation costs, especially as the number of shards grows.

In SP-Chain, to protect the security of cross-shard transactions with high efficiency, we propose the proof-assisted efficient cross-shard transaction processing mechanism.
Instead of heavy cross-shard proofs, special hardware reliance, complex two-phase commit exchanges, or full multi-shard consensus for each transaction, SP-Chain attaches compact cryptographic proofs (e.g., pruned Merkle proofs \cite{liu2021merkle}) to cross-shard transactions, allowing involved shards to independently verify transaction outcomes.
With such design, one pruned proof is attached to a batch of transactions to verify them together, reducing the overhead.
This design secures cross-shard transactions (even against adversarial senders or leaders) with high efficiency.
}


\vspace{3pt}
\noindent
\textbf{Issues from Intra-Shard Aspect.}
Existing sharding works reach consensus with limited efficiency.
A major reason is that most of them \cite{luu2016secure, kokoris2018omniledger, huang2020repchain, dang2019towards, zhang2020cycledger, hong2021pyramid, huang2022brokerchain, li2022jenga, liu2023flexible, li2023cochain} use BFT-typed consensus protocols suitable for scenarios with poor network conditions (i.e., partially synchronous or asynchronous).
Those protocols usually require multiple rounds of communication ($\geq$3) to reach consensus, slowing down the efficiency of reaching consensus. 
However, {\color{red}it is widely assumed that the network has reasonably good synchronization within each shard (i.e., bounded delays), as is common in such systems}
\cite{zamani2018rapidchain, huang2020repchain, zhang2020cycledger, li2020polyshard, kokoris2018omniledger}. 
In \cite{zamani2018rapidchain}, authors apply a BFT-typed consensus protocol that is suitable for good network condition, yet their protocol still require complex communication.
Authors in \cite{abraham2020sync} also propose a consensus protocol for good network condition, but their protocol requires view change to change the leader under malicious cases, reducing efficiency.
Some works \cite{wang2019monoxide, li2023lb} use Proof-of-Work (PoW) protocol as their intra-shard consensus. 
However, PoW-based protocol typically has low performance as it cannot guarantee instant finality, and it is easy to fork.
{\color{red}
Cherubim \cite{liu2024cherubim} introduces a quadruple-pipelined two-phase commit (4P-2PC) that parallelizes intra-shard consensus and cross-shard transaction processing to boost throughput. 
However, the scheme's complexity still incurs significant coordination overhead. 
}
In SP-Chain, we exploit the good and synchronous network condition in each shard and propose the two-phase concurrent voting consensus protocol.
It reduces the number of communication rounds in consensus (2 rounds) under synchronous network without compromising security, and {\color{red}parallelizes both the computation and communication steps that would otherwise be handled serially in the traditional consensus process}, thus significantly improving the consensus efficiency. 

In previous sharding systems, the leader is vulnerable to be attacked. 
The reason is that, in existing protocols \cite{luu2016secure, zamani2018rapidchain, kokoris2018omniledger, hong2021pyramid, li2022jenga, huang2022brokerchain}, a leader keeps producing blocks for a period of time and the rotation of the leader can be known in advance, leaving the chance for attackers to attack the leader.
Moreover, when the leader is found to be malicious or attacked, their consensus protocols perform a complex view change process to replace the leader, which is inefficient. 
{\color{red}
Despite the emergence of various new types of blockchain sharding systems in the last few years, most of them, however, also suffer from the aforementioned vulnerability of the leader to attacks as well as malicious leaders affecting performance \cite{li2023lb, li2023cochain, liu2023flexible, liu2024cherubim, qi2024lightcross, zhang2024efficient, huang2024brokerchain, xu2024x}.
For example, Cherubim \cite{liu2024cherubim} treats the underlying Byzantine consensus as a black box and does not specifically address leader fairness -- a malicious or monopolizing leader could still hinder performance or bias the process.
LightCross \cite{qi2024lightcross} does not modify the intra-shard consensus (it builds on an existing PBFT-style engine FISCO-BCOS), so issues of leader fairness or suboptimal consensus throughput per shard remain unaddressed.
ShardCon's \cite{zhang2024efficient} design prioritizes cross-shard consistency and formal security proofs, but it does not fundamentally enhance intra-shard consensus performance or address the potential performance bottleneck of a faulty leader before a view change triggers.
In SP-Chain, we design an unbiased distributed randomness generation scheme with low overhead, and propose an efficient and secure leader rotation mechanism that regularly and randomly rotates the shard leader, based on our consensus protocol and the distributed randomness.
Consequently, SP-Chain avoids the single-leader performance bottleneck and mitigates malicious leader behavior, a feature absent in most of the recent state-of-the-art blockchain sharding systems (e.g., \cite{liu2023flexible, liu2024cherubim, qi2024lightcross, zhang2024efficient, huang2024brokerchain, xu2024x, jiang2024sharon, jia2024estuary}).
}

\subsection{Distributed Randomness}

Distributed randomness is often used by blockchain to generate random groups or to elect leaders. 
However, existing distributed randomness generation methods either can be biased or involve high communication complexity.
Algorand \cite{gilad2017algorand} proposes to use the verifiable random function (VRF) \cite{micali1999verifiable} to randomly select committee members. 
However, the randomness (seed) used in VRF can be biased by the adversary. 
Elastico \cite{luu2016secure} uses PoW results to generate randomness, which can be biased by adversaries.
Ouroboros \cite{kiayias2017ouroboros} uses the publicly verifiable secret sharing (PVSS) scheme \cite{cascudo2017scrape} to generate the random seed for leader selection.
Omniledger \cite{kokoris2018omniledger} and Gosig \cite{li2020gosig} leverages VRF \cite{syta2017scalable} for its unbiased leader selection. 
Rapidchain exploits verifiable secret sharing (VSS) \cite{feldman1987practical} for cryptographic sortition. 
However, these schemes involve high communication complexity ($O(n^2)$) and thus not efficient.
In SP-Chain, nodes use unbiased signature information from already confirmed blocks to quickly generate distributed randomness. 
This mechanism needs no extra communication among nodes and thus allows efficient and secure production of distributed randomness for leader election.

{\color{red}
\subsection{Alternative Cross-Shard Transaction Processing Schemes}

Alternatively, cross-shard transaction execution can be offloaded to layer-2 mechanisms such as state channels \cite{negka2021blockchain} and Plasma chains \cite{poon2017plasma}. In a state-channel-based approach, participants lock assets on-chain and conduct numerous transfers off-chain, only settling the final state on-chain. This avoids most on-chain cross-shard communication but requires continuous client involvement -- users must remain online (or employ trusted "watchers") to respond to disputes or malicious counter-parties, and the method typically only supports a limited set of participants per channel. 
Plasma sidechains take a different route by anchoring sharded (or child-chain) transactions to a main chain via periodic commitments. Plasma designs can facilitate many off-chain transactions, but they introduce significant trust assumptions and latency: users depend on an honest Plasma operator and must engage in complex exit protocols to withdraw funds safely. 
For example, funds withdrawals from a Plasma chain are often delayed by a predetermined challenge period (on the order of one to two weeks) to allow fraud proofs, and users need to actively monitor the system to ensure the security of their assets. 
These limitations -- high client burden, delayed finality, and reliance on external trusted parties or timely responses -- make purely off-chain approaches less desirable for cross-shard verification in an open system. 

In contrast, SP-Chain addresses the above limitations within its on-chain sharding design. 
The cross-shard transaction mechanism of SP-Chain is integrated into the protocol so that no special action is required from clients beyond initiating the transaction. 
Unlike OmniLedger's client-dependent cross-shard commits or state channels that assume 100\% availability of participants, SP-Chain achieves atomic cross-shard confirmation without user intervention. 
This is enabled by a proof-assisted verification: each cross-shard transaction carries a cryptographic proof (using pruned Merkle proofs) that can be batched and verified by the destination shard's validators, eliminating any need to trust a separate intermediary (as in Plasma) or to exchange multiple round-trip messages off-chain. 
By enforcing cross-shard validity through the consensus of involved shards, SP-Chain maintains security under standard BFT assumptions and avoids introducing new trust dependencies. 
In summary, SP-Chain's approach preserves the trustless security of on-chain sharding while minimizing overhead, effectively overcoming the client participation requirements and security vulnerabilities of state-channel and Plasma-based solutions.
}

%% file: model.tex
\section{System and Threat Model}
\label{sec:model}

\subsection{System Model}

SP-Chain proceeds in epochs. 
There are multiple slots $t$ in each epoch $e$.
We assume there are $n$ nodes in the network for each epoch (noting that $n$ might be changing as epoch changes).
Each node is given a public/secret key pair $(PK,SK)$ (e.g., via a {\color{red}Public-Key Infrastructure (PKI)} \cite{zhang2020cycledger}). All nodes are partitioned into $m$ shards (a.k.a.
committees). 
Thus, there are $k=n/m$ nodes (a.k.a. members) in each shard, including one leader. 


Our network model is similar to many previous works \cite{zamani2018rapidchain, huang2020repchain, zhang2020cycledger}. 
Specifically, the authenticity of all messages disseminated in the network is protected by the signature of the sender.
The connections between honest nodes are well connected. 
Like many sharding-related studies \cite{zamani2018rapidchain, huang2020repchain, zhang2020cycledger, li2020polyshard}, we use a synchronous gossip protocol \cite{karp2000randomized} to transmit messages across the network.
This means that, within a pre-known, fixed amount of time \ensuremath{\Delta}, any message that is sent or forwarded by an honest node will be delivered to all honest nodes, i.e., the communication network is \emph{synchronous} within each shard.
To address the issue of poor responsiveness existing in any synchronous consensus and achieve long-term responsiveness, we require each shard to agree on a new \ensuremath{\Delta} for about once a week, which is a similar approach to existing works \cite{zamani2018rapidchain}.
In addition to the intra-shard consensus, the rest of SP-Chain is built on the assumption of a partial-synchronous network.
Without loss of generality and similar to many other blockchain systems \cite{zamani2018rapidchain, kokoris2018omniledger, luu2016secure, huang2020repchain, zhang2020cycledger, wang2019monoxide}, all nodes that participated in our system have equivalent and \emph{enough computational resources. }

SP-Chain adopts the account model to represent the state of the blockchain, where each account has its own states.
The states of one account are maintained by one certain shard, for computation and storage scalability.
Which shard an account's state should be stored by is determined by its address. 
The account address is mapped to a shard based on the output of a random oracle (e.g., the remainder of the account address divided by the number of shards).
When an account initiates a transaction, that transaction is routed to the corresponding shard based on the address of its sender account.
How to design smarter allocation mechanisms for accounts and transactions (e.g., \cite{chen2021user, ren2021evaluating, huang2022brokerchain, li2023lb, qi2024lightcross}) is orthogonal to this work and will therefore be discussed in our future work.

\subsection{Threat Model}
\label{sec:threat}

We build a similar threat model as previous works do \cite{zamani2018rapidchain, huang2020repchain, zhang2020cycledger}. 
In our model, there exists a Byzantine adversary who can take control of $<1/3$ fraction of the \emph{total nodes}. 
Similar to the previous works, the communication channel is synchronous in one shard. 
Therefore, \emph{each shard} can achieve an optimal fault resiliency of 1/2.
Corrupted (Byzantine) nodes may collude and behave arbitrarily, such as generating and sending invalid messages (transaction manipulation), sending messages to different nodes with different values (equivocation), or not sending any or all of the messages (silence attack).
Other nodes besides the above are called honest nodes, they will always obey the protocol and do not do anything beyond what is specified.

We assume the adversary is \emph{mildly-adaptive}, which is a similar assumption to most existing blockchain sharding works, meaning that the adversary can only corrupt a fixed set of nodes at the beginning of each epoch (e.g., one day) and the set of corrupted nodes remains unchanged within an epoch.
Moreover, \emph{we allow the adversaries to have stronger attack ability} than previous studies, who can launch target attack on leaders.
For example, an adversary can bribe a leader and change the leader to a Byzantine node, or it can launch DDoS attacks against a leader so the leader's message cannot be transmitted to others.
Also, all nodes can access to a collision-resistant external random oracle, similar to other works \cite{zamani2018rapidchain, zhang2020cycledger}.

{\color{red}
To mitigate \emph{Sybil attacks}, similar to existing works \cite{kokoris2018omniledger, zamani2018rapidchain, huang2020repchain, li2023lb, Harmony}, each participating node must obtain a Sybil-resistant identity through an offline proof-of-work puzzle, or proof-of-stake, or PKI-based identity issuance process. 
This requirement makes the generation of new identities expensive and verifiable, preventing a malicious adversary from cheaply acquiring arbitrarily many node identities. Thus, the adversary is effectively limited to controlling at most the assumed fraction (<1/3) of all nodes. Moreover, at each epoch's shard reconfiguration, node identities are uniformly re-distributed across shards and validated (see Section \ref{sec:reconfigure}), ensuring that no shard can be overwhelmed by colluding Sybil nodes. These measures, built into the protocol, collectively enforce Sybil-resistance under our adversary model.
}

\section{System Overview}

\begin{figure}[t]
\centering
\includegraphics[scale=0.42]{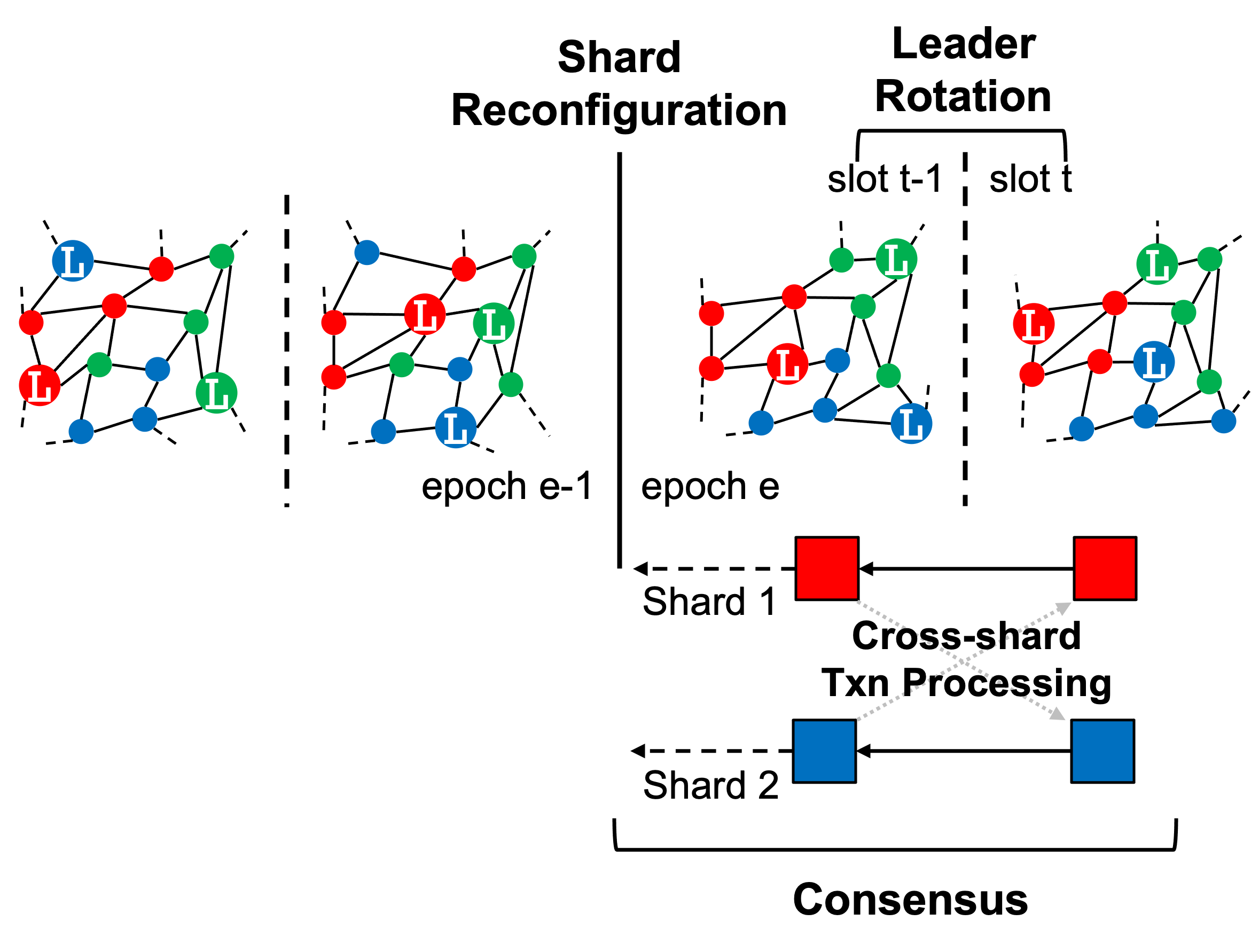}
\vspace{-18pt}
\caption{Overview of SP-Chain.}
\label{fig:model}
\end{figure}

SP-Chain consists of four main components, leader rotation, intra-shard consensus, cross-shard transaction processing, and shard reconfiguration, shown in Figure \ref{fig:model}. 
Our protocol selects a new leader in each slot to propose one block in each shard based on unbiased randomness. 
Then, the intra-shard consensus is executed.
When the consensus is reached, the (cross-shard) transactions are sent and executed. 
Each epoch consists of multiple slots followed by a reconfiguration phase, during which the shard members are reshuffled.
We now explain each component and the design intuition in more detail.

\vspace{3pt}
\noindent\textbf{Leader Rotation.} 
{\color{red}To prevent attacks on the leaders and maintain high performance when leaders are malicious or attacked, we propose a leader rotation scheme.
The leader rotation in SP-Chain is a mechanism by which a new block proposer is chosen every block-producing cycle (i.e., one slot).
}
Specifically, the leaders are changed frequently and randomly to prevent the attackers from knowing the leaders and launching targeted attacks.
In each slot, each shard elects a new leader among the nodes.
To ensure the security of the leader rotation process, the leader shall be elected based on an unbiased randomness retrieved from the confirmed block. 
Our distributed randomness generation scheme guarantees high efficiency when electing the new leader, as no extra communications are required.
More importantly, as the leaders are rotated automatically in each slot (no matter it is malicious or honest), the view change process is eliminated, achieving high efficiency even under attacks.

\vspace{3pt}
\noindent\textbf{Intra-Shard Consensus.} 
To boost the intra-shard consensus performance, we propose the two-phase concurrent voting consensus protocol.
We leverage the features of good network synchronization within each shard, and propose an efficient synchronous consensus protocol with less rounds of communication.
More importantly, {\color{red}we make both computation and communication steps run in parallel (which are handled sequentially in previous protocols)}, and propose the concurrent voting protocol to ensure high block generation and confirmation speed, hence improving system performance.

\vspace{3pt}
\noindent\textbf{Cross-Shard Transactions.} 
After a block in a shard is confirmed, the cross-shard transactions are sent to corresponding shards.
We should efficiently resist the malicious behaviors of leaders in cross-shard transaction processing.
Therefore, we propose the proof-assisted cross-shard transaction processing scheme.
In this mechanism, the transactions in a block are divided into batches according to the different shards to which they are sent.
One pruned proof is generated and attached for one batch of transactions to reduce the overhead.
The shard who receives the cross-shard transactions then verifies them based on the proof and packs them into the block. 

\vspace{3pt}
\noindent\textbf{Shard Reconfiguration.} 
Reconfiguration happens at the end of each epoch. 
During the reconfiguration phase, all the shards reshuffle their shard members.
SP-Chain applies the Cuckoo rule \cite{sen2012commensal, zamani2018rapidchain} for reconfiguration to allow the shard's nodes to change. 





%% file: design.tex
\section{Protocol Design}
\label{sec:protocol}

This section presents the detailed design of SP-Chain. 
We first describe our two-phase concurrent voting intra-shard consensus in Section \ref{sec:consensus}. 
Based on the proposed consensus protocol, we describe the leader rotation mechanism in Section \ref{sec:leader}. 
Next, we describe how cross-shard transactions are efficiently and securely processed in Section \ref{sec:cross-shard}.
Finally, finish this section by briefly describing the shard reconfiguration in Section \ref{sec:reconfigure}.

\begin{figure}[t]
\centering
\includegraphics[scale=0.33]{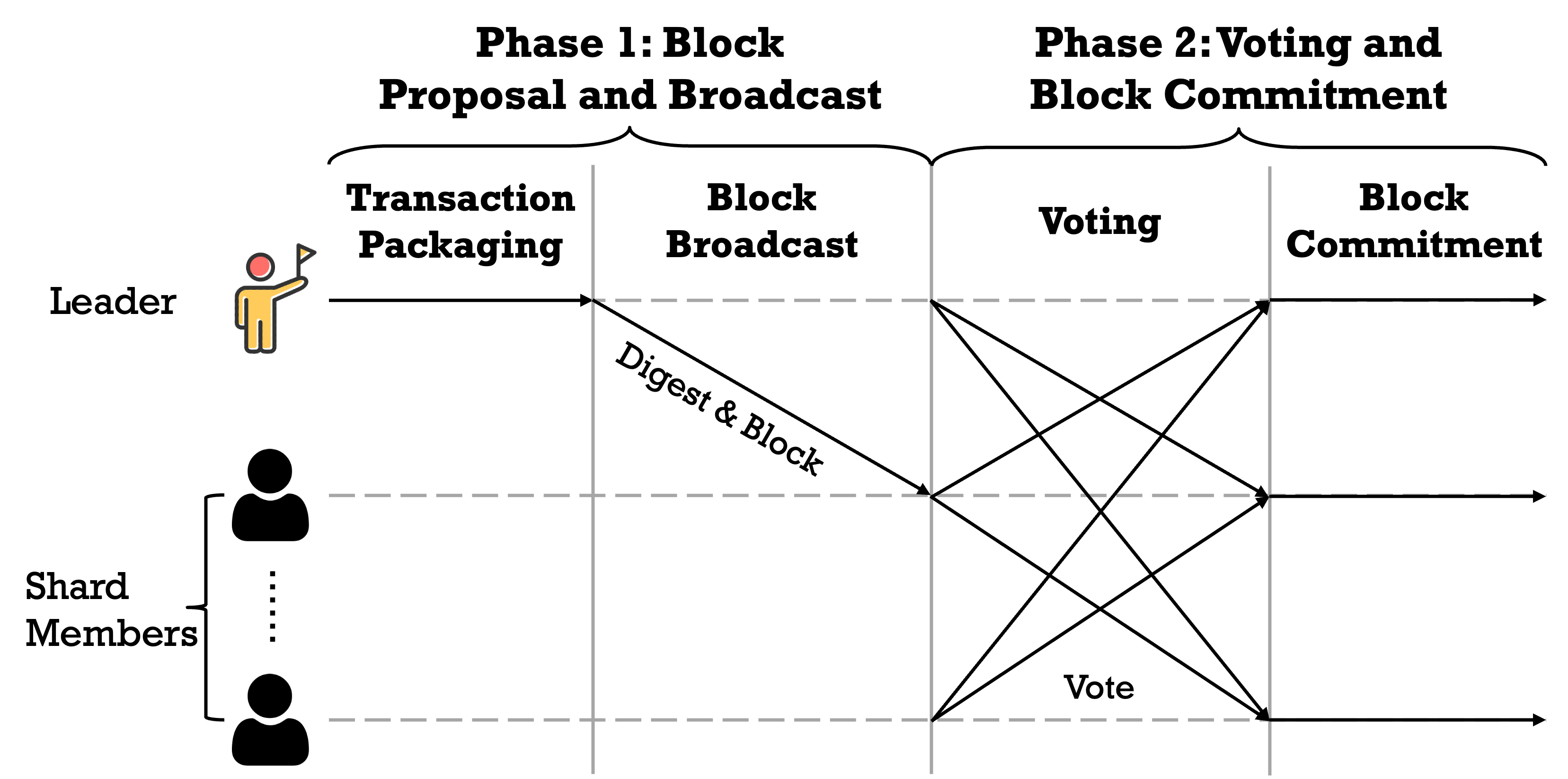}
\vspace{-12pt}
\caption{Overview of two-phase voting intra-shard consensus.}
\label{fig:intra-shard}
\end{figure}

\subsection{Intra-Shard Consensus}
\label{sec:consensus}

Our intra-shard consensus mechanism has two major design points: 
1) Based on the feature that each shard contains a small number of nodes and has a good network synchronization rate, we design an efficient, BFT-typed synchronous voting consensus protocol with two-phase communication;
2) Based on the two-phase voting, we propose the concurrent voting scheme that converts the serially handled computation and communication steps in traditional consensus into parallel.

{\color{red}
\subsubsection{Two-Phase Voting}

Each consensus round in a shard consists of two phases under a synchronous network model: phase 1 -- block proposal and broadcast, followed by phase 2 -- voting and block commitment. Figure \ref{fig:intra-shard} illustrates the workflow for a single round. By leveraging network synchrony (known message delay bounds) and a small committee size per shard, this design reduces the communication rounds from the usual three (pre-prepare, prepare, commit in PBFT-style protocols) to two, while still tolerating up to 50\% faulty nodes in each shard (which corresponds to the standard <1/3 byzantine fault tolerance across the network).

\vspace{3pt}
\noindent
\textbf{Phase 1: Block Proposal and Broadcast. }
At the start of a slot, the elected leader assembles pending transactions into a new block and broadcasts the block along with a short digest of the block to all other shard members. The block digest includes essential metadata (e.g. block height, slot number, leader ID) and a hash of the block, allowing members to quickly perform a pre-check before the full block arrives. Because the digest is small, it propagates faster: $\Delta_d$ denotes the maximum network delay for the digest, whereas $\Delta_b$ is the maximum delay for the full block. Upon receiving the digest (at some time $\delta \leq \Delta_d$ from the start of broadcast), each member starts a local timer and waits for the full block. No global clock sync is required -- each node uses the digest's arrival as a reference point for timing. By time $\Delta_b$, every honest member should have received the complete block from the leader. (If the block fails to arrive in time or is invalid, the protocol will reject that leader's proposal and move to the next slot.) 

\vspace{3pt}
\noindent
\textbf{Phase 2: Voting and Block Commitment. }
Once a member's local timer reaches $\Delta_b$ after digest reception, it proceeds to validate the received block and then broadcasts its vote on the block to the shard. A vote contains the validator's signature, the current slot number, and an indication of whether the block is considered valid. We let $\Delta_v$ be the maximum network delay for vote messages. Because different validators may start their timers at slightly different moments (up to $\Delta_d$ apart due to network latency of the digest), each validator waits an additional $\Delta_d + \Delta_v$ after sending its own vote to ensure receipt of votes from all other members. By the end of this voting phase, every member will have collected votes from a majority of the committee (when the leader and network behaved correctly). If more than half of the shard's nodes voted to approve the block, the block is committed to the shard's blockchain. Each node then officially adds the new block to its ledger. If the block fails to gather a majority of yes-votes (or if the leader's proposal was never received), the block is rejected and the shard simply moves on to the next slot without committing it.

\vspace{3pt}
\noindent\textit{Remarks.} 
There are usually 3 rounds of communication (e.g., pre-prepare, prepare, commit) in previous consensus protocols \cite{luu2016secure, kokoris2018omniledger, huang2020repchain, zhang2020cycledger} under partial-synchronous network.
In our proposed protocol, there are 2 rounds of communication.
The first round is block broadcast, which is similar to the pre-prepare phase in previous consensus.
The second round is voting, which can be seen as the compaction of the prepare and commit phase.
The two-phase voting consensus protocol exploits the features of the synchronous network in each shard, such as the guaranteed delay upper bound and the high connectivity between honest nodes. 
In this synchronous network context, the proposed protocol reduces the communication overhead without compromising security.
More importantly, our design leverages the fact that each shard consists of a small number of nodes. 
This feature keeps $\Delta_{d}$, $\Delta_{b}$ and $\Delta_{v}$ at a small value, which shortens the interval of each slot and improves the protocol efficiency.

\begin{figure}[t]
\centering
\includegraphics[scale=0.33]{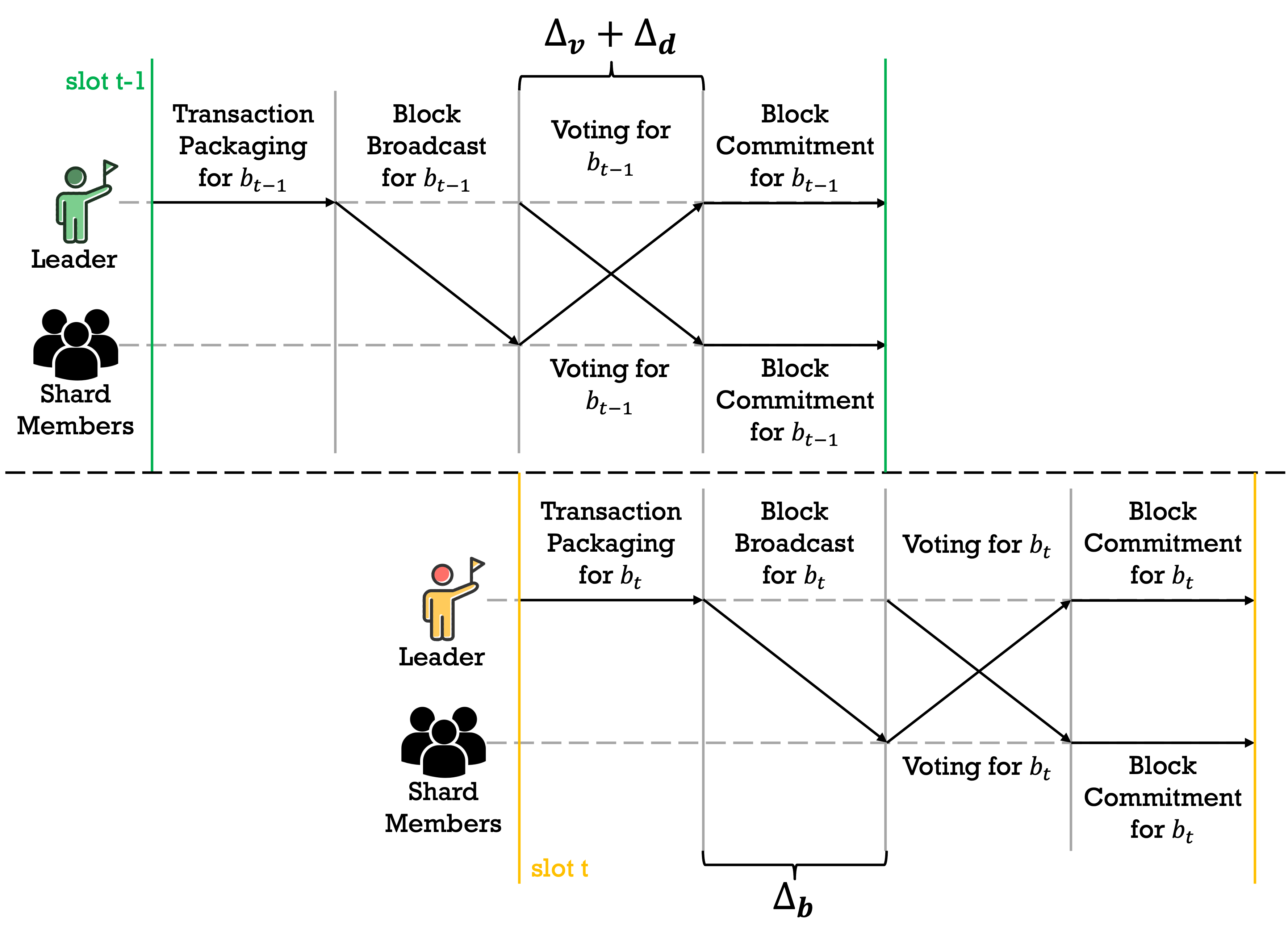}
\vspace{-21pt}
\caption{Overview of pipelined concurrent voting.}
\label{fig:consensus}
\end{figure}

\subsubsection{Concurrent Voting}

To maximize throughput, SP-Chain \emph{\textbf{pipelines}} the above two phases across consecutive slots, shown in Figure \ref{fig:consensus}. This means nodes can work on different phases of different blocks in parallel. For example, while the shard members are busy voting on the block from slot $t-1$ (Phase 2 for block $b_{t-1}$), the new leader for slot $t$ can concurrently begin packaging transactions for the next block $b_t$ (the start of Phase 1 for that round). Likewise, when the leader of slot $t$ later broadcasts block $b_t$, the members are simultaneously committing block $b_{t-1}$. This two-slot overlap ensures that there is almost no idle time: \emph{computation (block assembly and insertion) is done in parallel with communication (broadcasting and voting) in a continuous pipeline}. 


\vspace{3pt}
\noindent\textbf{Additional Challenges and Solutions.}
The concurrent voting brings additional security and efficiency challenges. 
Details are described as follows.

First, the concurrent voting enables the leader to pack the transactions into block $b_{t}$ when the members are voting for block $b_{t-1}$.
However, in concurrent voting, the leader when generating block $b_{t}$ cannot decide which block should $b_{t}$ be chained after, as the voting for the last block is not over yet. 
Therefore, after packing transactions, the leader should wait for the voting finish to finally seal the block $b_{t}$. 
The voting contains the voter's signature, current slot number, and the hash of the latest valid block the voter thinks.
When the voting phase is over, the leader decides which block should $b_{t}$ be chained after, according to the voting results (e.g., when more than $k/2$ of the members think $b_{t-1}$ is valid, then $b_{t}$ is chained after $b_{t-1}$).
The leader then constructs the block header and finish the transaction packaging.

Second, if block $b_{t-1}$ is invalid, the transactions in block $b_{t}$ might need to be re-packed in concurrent voting, which damage the system performance. 
To address this, the leader's transaction packaging for block $b_{t}$ needs to be executed after it receives the broadcast of block $b_{t-1}$.
After the leader for generating block $b_{t}$ receives the broadcast for $b_{t-1}$, it first verifies $b_{t-1}$ and packs transactions which are mutually exclusive to the transactions in $b_{t-1}$ (e.g., transactions sent by different accounts). 
In this way, the block $b_{t}$ can be generated successfully no matter the block $b_{t-1}$ is valid or not. 

\vspace{3pt}
\noindent\textit{Remarks.}
Our concurrent voting is different from previous pipelined consensus protocols \cite{li2020gosig, yin2019hotstuff}, where they focus on the parallelization of the communication.
In concurrent voting, we decouple the consensus process in a more fine-grained way, where both communication and computation are parallelized.
Also noting that the computation power is usually assumed to be sufficient. 
Therefore, the network is the bottleneck, rather than computation. 
Hence, in practice, the transaction package and block commitment (computation) take less time than voting and block broadcasting, respectively.
Finally, since each node's clock is not synchronized, a node may receive a vote (or digest) before its waiting time $\Delta_{d}+\Delta_{v}$ (or $\Delta_{d}$) is over. 
In this case, the node verifies the received message while waiting for the waiting time to end.

\vspace{3pt}
\noindent
\textbf{Theoretical Basis for Parameter Choices ($\Delta_{d}$, $\Delta_{b}$, $\Delta_{v}$).}
Under the synchronous network assumption (i.e., every message is delivered within a known maximum delay $\Delta$), we choose the intra-shard communication timeouts $\Delta_d$, $\Delta_b$, and $\Delta_v$ to be small multiples of this bound for each consensus phase (see Section \ref{sec:ExpSetup} for exact values). 
The theoretical rationale is that allocating on the order of one full worst-case network delay per sub-phase provides enough slack to absorb worst-case delay variance while ensuring all honest nodes receive the required digests, blocks, and votes before progressing. 
By waiting up to a couple of $\Delta$ intervals in each phase (e.g., a slightly longer window for the larger block broadcast, and around one $\Delta$ for smaller digest and vote messages), the protocol tolerates maximum network latency differences so that even the slowest honest participant's message arrives in time. 
At the same time, keeping these delays a bit longer than $\Delta$ preserves consensus responsiveness, as nodes move to the next step as soon as the expected worst-case propagation time elapses, avoiding any unnecessary idle waiting. This careful setting of $\Delta_d$, $\Delta_b$, and $\Delta_v$ thus upholds consensus safety under the synchronous model (no node decides prematurely) while minimizing latency overhead, maintaining both reliability and efficiency in reaching agreement.
}

\subsection{Leader Rotation}
\label{sec:leader}

Based on the concurrent voting consensus protocol, we now propose the random leader rotation mechanism to determine each block's producer securely and efficiently. 
When a node enters a new slot,
it will judge whether to be the new leader through distributed randomness generation. 
Each leader is responsible for proposing one new block. 
To prevent malicious nodes from biasing the result of leader rotation, the choice of randomness is critical. 
Therefore, we propose an \emph{unbiased distributed randomness generation scheme} that can ensure the security of leader rotation.

\vspace{3pt}
\noindent\textbf{Chain-Based Randomness Generation.} 
At the beginning of each slot, a new random number is calculated to elect the leader of that slot. 
Specifically, when a node starts slot $t$, it extracts the signature information from the latest confirmed block (e.g., $b_{t-2}$, as the consensus is not reached for $b_{t-1}$ due to concurrent voting). 
Each node uses the signature information and the current slot number $t$ as a seed, input the seed to a publicly known pseudo-random number generation function (i.e., the random oracle mentioned in Section \ref{sec:threat}). 
This function will uniformly map the value of the seed to one of the nodes. 
The selected person is the leader in slot $t$ and is responsible for generating block $b_{t}$.

\vspace{3pt}
\noindent\textbf{Choice of Signature Information.} 
The most crucial point in the above process is the choice of signature information. 
When the selected signature information is not biased, it can be ensured that the leader election is unbiased. 
For this reason, we design that when $l_t$ (leader of slot $t$) generates a block, it will sign the current slot number $t$ and leave the signature information $SIG_{l_t}(t)$ in the block header.
In this way, each member can use the signature from the latest confirmed block (e.g., $SIG_{l_{t-2}}(t-2)$) as the seed to calculate the new leader. 
For example:

\vspace{-9pt}
\begin{equation}
l_{t}=H(SIG_{l_{t-2}}(t-2),t),
\end{equation}
where $H(\cdot)$ is the random oracle that uniformly maps the input to one of the leader candidates.

\subsection{Cross-Shard Transactions}
\label{sec:cross-shard}

The processing of cross-shard transactions should ensure safety and efficiency simultaneously. 
Specifically, since each shard does not know other shards' state, a shard that receives cross-shard transactions (a.k.a. destination shard) cannot directly verify whether the received cross-shard transactions are confirmed in the source shard (shard that sends cross-shard transactions). 
A malicious leader can therefore \emph{send arbitrary cross-shard transactions or generate dummy signatures} to deceive the destination shards \cite{zhang2020cycledger}. 
To prevent such problems, we design a low-overhead, proof-assisted cross-shard transaction processing scheme to achieve the purpose of ensuring security while maintaining high efficiency.

\begin{figure}[t]
\centering
\includegraphics[scale=0.36]{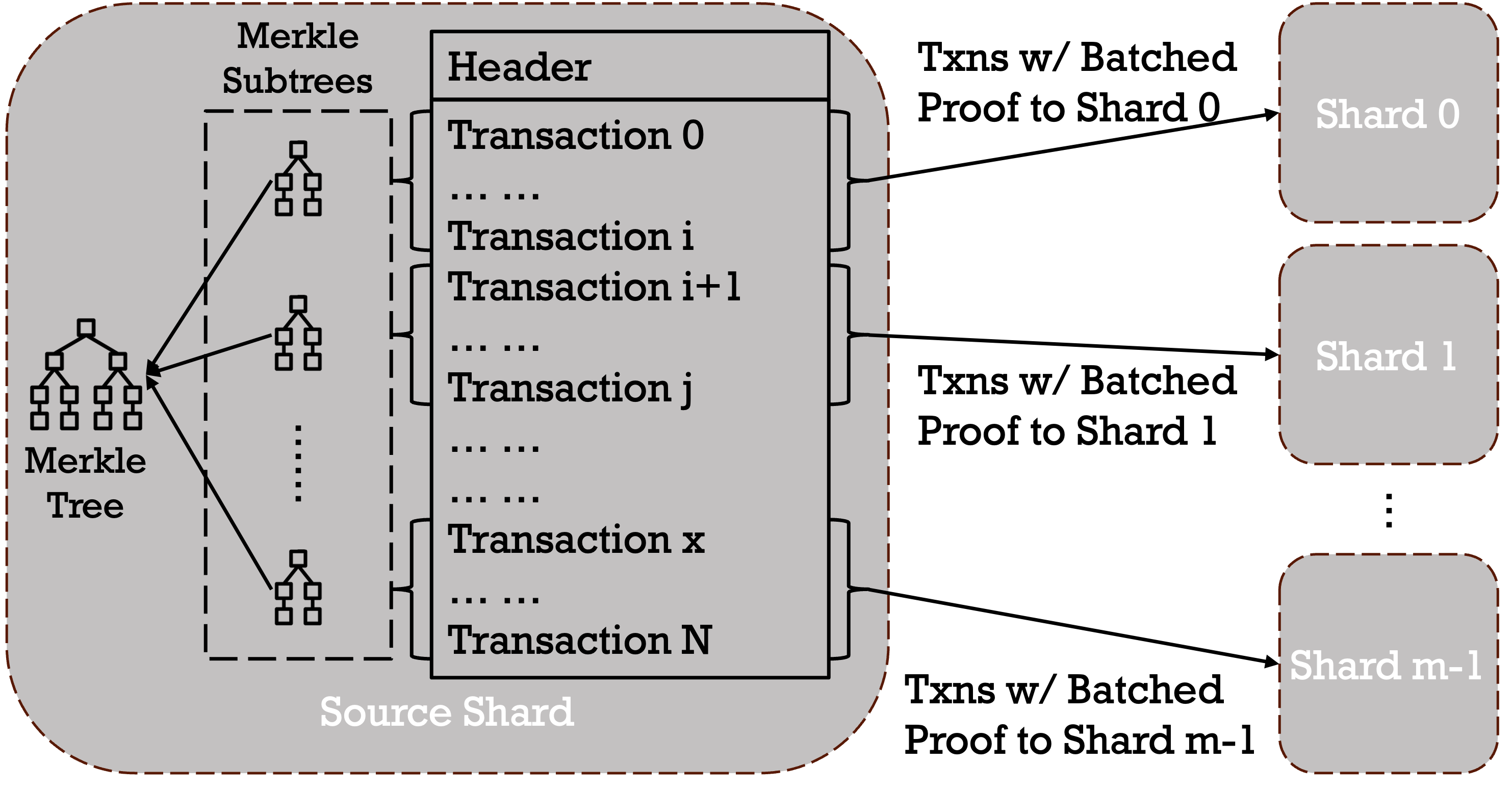}
\vspace{-12pt}
\caption{Cross-shard transactions processing in a block.}
\label{fig:block}
\end{figure}

{\color{red}
We leverage the Merkle Tree (i.e., a cryptographic tree structure where each leaf is a hash of data, and each non-leaf node is the hash of its children) to verify the correctness of a transaction.
A straightforward idea is to attach a Merkle proof (i.e., a way to prove that a leaf is part of a Merkle tree by showing the necessary sibling hashes)
} 
to each cross-shard transaction so that the destination shard can verify the validity of the cross-shard transaction, but this approach introduces lots of overhead. 
To reduce the overhead, \emph{our main idea} is that, for the transactions in a block, we \emph{arrange them and construct Merkle Tree according to different shards}.
Based on the constructed Merkle Tree, the \emph{transactions sent to the same shard are attached with one pruned Merkle proof together}. 
One such proof can be used to verify this batch of transactions simultaneously.
An illustration for processing cross-shard transactions is shown in Figure \ref{fig:block}.

\vspace{3pt}
\noindent\textbf{Constructing Merkle (Sub)trees According to Shards.} 
The transactions packed into a block are no longer randomly arranged and constructed into a Merkle Tree as traditional approaches. 
In SP-Chain, transactions sent to the same shard will be sorted first according to their transaction hash values, as shown in Figure \ref{fig:block}. 
The \emph{sorted transactions sent to the same shard will then be constructed into a Merkle Subtree}. 
Different Merkle Subtrees (representing transactions sent to different shards) will be merged into a complete Merkle Tree, including transactions sent to all shards.
The Merkle root of the complete Merkle Tree will be written into the block header and verified by members during consensus.

\vspace{3pt}
\noindent\textbf{Sending of Cross-Shard Transactions.} 
After a block is confirmed by consensus, the leader producing the block will send the cross-shard transactions contained in the block to the corresponding shards. 
While sending cross-shard transactions, the leader will broadcast the block header to all other shards.
To enable cross-shard transactions to be verified by the destination shard, the leader also needs to broadcast the roots of all Merkle Subtrees to the network.
In our design, cross-shard transactions sent to the same shard will be sent in batch.
Moreover, the Kademlia routing algorithm \cite{maymounkov2002kademlia} is used for the routing of cross-shard transactions.

\vspace{3pt}
\noindent\textbf{Receipt and Verification of Cross-Shard Transactions.} 
After receiving the messages mentioned above, the destination shard 
reconstructs the corresponding Merkle Subtree root, and then reconstructs the Merkle Tree root based on other received Merkle Subtree roots. 
After the reconstruction, any node in the destination shard can judge whether the received transactions are modified by comparing the reconstructed Merkle Tree root with the Merkle Tree root in the signed block header. 
To prevent the leader from forging shard members and signatures, a shard member table (see Section \ref{sec:reconfigure} for details) is maintained by each node. 
The table contains the valid public keys of all nodes in all shards. 
In this way, after receiving the header, the destination shard can check whether there is an illegal member's signature by comparing the member table.

\subsection{Shard Reconfiguration}
\label{sec:reconfigure}

The main components of our shard reconfiguration are similar to RapidChain \cite{zamani2018rapidchain}, which includes: 1) Offline PoW to prevent Sybil attacks;
2) Epoch randomness generation; 3) Committee reconfiguration; 4) node fast initialization after joining the committee.
The main difference between our design and theirs lies in the epoch randomness generation. 
The epoch randomness is used to solve the offline PoW and reshuffle shard members.
In SP-Chain, for efficiency and security, we use our chain-based randomness to generate epoch randomness. 
Specifically, since the block headers are broadcasted (Section \ref{sec:cross-shard}), the reference committee (a.k.a. beacon chain) \cite{zamani2018rapidchain, Harmony} can collect the seed information in all confirmed blocks in the last epoch.
The reference committee then XOR the seeds to obtain a new epoch seed and use it to generate the epoch random number.
Each new node can request the randomness of this epoch as a fresh PoW puzzle. 
Additionally, during shard reconfiguration, node change information (join or leave) will be broadcast. 
According to the node change information, each shard generates a state block containing the shard member table (of all shards).

%% file: analysis.tex
\section{Security and Performance Analysis}

We first analyze the system failure probability during each epoch.
Under negligible system failure probability, we then analyze the security for our main components and discuss their overhead.

\subsection{Epoch Security}
We first calculate the failure probability of each epoch. 
Similar to previous works \cite{dang2019towards, zamani2018rapidchain, kokoris2018omniledger, luu2016secure}, we use the hypergeometric distribution for calculation. 
In particular, let $X$ be a random variable representing the number of Byzantine nodes assigned to a shard of size $k=n/m$, given the overall network size of $n$ nodes among which up to $F$ nodes are Byzantine. The failure probability for the system in each epoch is at most:

\vspace{-9pt}
\begin{equation}
Pr[X\geq\lfloor k/2\rfloor]=m\cdot\stackrel[x=\lfloor k/2\rfloor]{k}{\sum}\frac{\binom{F}{x}\binom{n-F}{k-x}}{\binom{n}{k}}.
\label{equ:failure}
\end{equation}

\noindent\textbf{Ensuring Negligible Failure Probability.}
We should carefully choose the shard size to bound the failure probability of the system to be negligible.
As will be explained in Section \ref{sec:ExpSetup}, our choice of shard size can limit the failure probability to be less than $2^{-20}\approx9\cdot10^{-7}$ (time-to-failure of more than 4580 years for one-day epoch), which is a wide-adapted probability threshold \cite{dang2019towards, zamani2018rapidchain}.

\vspace{3pt}
Under negligible epoch failure probability, we next analyze the security and performance of our system.

{\color{red}
\subsection{Analysis of Randomness Generation and Leader Rotation}
\label{sec:ConsensusSecurity}

We formalize the security of our chain-based distributed randomness scheme for leader rotation. In particular, we prove that the scheme's output is unbiasable (fair) and unpredictable under standard cryptographic assumptions, meaning no adversary can manipulate or guess the leader selection beyond a negligible probability. We also note that the randomness is publicly verifiable and the generation process is scalable (incurring minimal overhead) \cite{syta2017scalable}.

\begin{definition}[Unbiasability/Fairness]
    A randomness generation scheme is unbiasable if no coalition of Byzantine (malicious) nodes, up to a threshold $f$, can influence the distribution of the random output. In the context of leader rotation among $k$ shard members, unbiasability means each member has an equal probability $1/k$ of being selected as leader (deviating at most negligibly due to adversarial actions).
\end{definition}

\begin{definition}[Unpredictability]
    A scheme's output is unpredictable if, prior to its generation, no computationally bounded adversary can guess the output with more than negligible advantage over random chance. For leader rotation, this means that until the randomness is revealed (by the confirmation of the current block), the identity of the next leader cannot be predicted except with probability close to $1/k$.
\end{definition}

\begin{thm}[Fairness of Leader Selection]
    Under the assumption of an honest majority in each shard (at most $f < k/2$ Byzantine nodes in a shard of size $k$) and an existentially unforgeable digital signature scheme, the leader selection in each slot of SP-Chain is unbiased. In other words, the randomness derived from the previous block's signature yields a uniform random leader except with negligible probability.
\end{thm}

\begin{proof}
    The leader for each new slot is determined by applying a public random oracle $H(\cdot)$ (modeled as a cryptographic hash function) to the prior confirmed block's signature (and slot number). Because the slot number and block contents are fixed once a block is finalized, the input to $H$ is fixed and outside an adversary's control at the time of leader election. A malicious leader cannot bias this process: given a unique identity and key pair per node (enforced by Sybil-resistant identity issuance and periodic reconfiguration of shard membership), a leader can produce at most one valid signature on a given block. Any attempt to produce an alternate seed (e.g., by forging another node's signature or using an illegitimate identity) would require breaking the signature scheme's unforgeability or introducing a fake identity, both of which are assumed infeasible (negligible probability). Furthermore, since $H$ is random-oracle unpredictable, the output is uniformly distributed over the space of possible leader identifiers. No biased influence can be exerted by the adversary beyond choosing not to sign or withholding the block (which would simply forfeit their turn). Thus, each eligible node has equal chance to be chosen by the hash, and no adversary (up to $f$ nodes) can skew this distribution except with negligible probability (e.g., by controlling a proportion of identities, which is bounded by $f/k$ for each shard). This establishes unbiasability of the leader rotation mechanism.
\end{proof}

\begin{thm}[Unpredictability of Leader Selection]
    The signature-based randomness used for leader rotation is unpredictable until the moment of its use. Formally, even a Byzantine adversary (with up to $f$ corrupted nodes per shard) cannot obtain any information about the next slot's leader prior to the decision of the lastet confirmed block, except with negligible probability.
\end{thm}

\begin{proof}
    At the start of a slot $t$, the new leader is determined from the latest confirmed block's signature (from slot $t-2$). By the properties of the hash function $H$, the mapping from a valid signature $\sigma$ to an output leader index is computationally indistinguishable from random. No party learns $\sigma$ (hence the next leader) until the block at $t-2$ is confirmed (it is confirmed at the beginning of $t-1$). In a synchronous network, all honest nodes finalize the block (and observe $\sigma$) nearly simultaneously within a known bound $\Delta$. A malicious leader at $t-2$ gains no useful advantage: they cannot selectively choose $\sigma$ to influence $H(\sigma||t)$ (by Theorem 1, $\sigma$ is essentially fixed by the correct signing process), and they learn the next leader only moments before others (at most on the order of the network delay $\Delta$). Thus, prior to the block confirmation at $t-2$, the probability of any party predicting the next leader is at most $1/k$ plus a negligible factor. This satisfies the unpredictability property.
\end{proof}

\noindent
\textbf{Verifiability and Scalability. }
The randomness generation is publicly verifiable -- all inputs (the previous block hash/signature, public slot number, and the hash function) are publicly known, so any node or third party can independently recompute and verify the leader selection outcome. 
Moreover, the scheme is scalable: deriving randomness from existing blockchain data avoids any extra communication rounds. Each node simply performs a local hash computation to determine the leader, incurring $O(1)$ communication overhead (in contrast to heavier protocols like VSS-based beacons). This means our leader rotation mechanism adds virtually no networking cost while providing cryptographically strong randomness.

\subsection{Analysis of Intra-Shard Consensus}
\label{subsec:intra_security}

We now rigorously define and prove the safety and liveness properties of the intra-shard consensus protocol (the pipelined two-phase concurrent voting combined with periodic leader rotation). We assume a synchronous network model with known maximum message delay $\Delta$, and that each shard has an honest majority ($f < k/2$ Byzantine nodes out of $k$). Under these conditions, we show that the consensus protocol maintains consistency (all honest nodes agree on the same shard state) and makes progress (new blocks eventually commit).

\begin{definition}[Safety -- Consistency]
    A consensus protocol satisfies safety (or consistency) if no two honest nodes ever decide different blocks in the same round/slot. In a blockchain shard, this means that at the end of each slot, all honest shard members have committed the same block (or know that no block was committed in that slot). Equivalently, there are no "forks" within a shard's ledger visible to honest participants.
\end{definition}

\begin{definition}[Liveness]
    Liveness means that the protocol continues to make progress despite adversarial interference. For the shard consensus, liveness requires that any valid transaction submitted will eventually be included in a committed block. More concretely, in every epoch (or within a bounded number of slots), if the current leader is malicious or fails, the protocol will still eventually commit a new block under a future leader. This ensures that the shard's chain grows and honest transactions are not starved indefinitely.
\end{definition}

\begin{thm}[Safety and Liveness of Intra-Shard Consensus]
    Under the synchronous network and honest-majority assumptions, the two-phase concurrent voting protocol with automatic leader rotation achieves both safety and liveness: (i) Safety: All honest nodes in a shard agree on the same block each slot (no conflicting commits occur). (ii) Liveness: The consensus will successfully commit new blocks in a timely manner, even in the presence of malicious leaders, as long as at least one leader in a succession of slots is honest (which is guaranteed by the rotation and majority assumption).
\end{thm}

\begin{proof}
    \emph{\textbf{Safety}}: We first consider a single slot with a designated leader. When an honest leader proposes a block, all honest replicas will receive the block digest (hash) within time $\Delta$ (by synchrony). Due to the two-phase voting design, an honest node waits a fixed pre-vote period (e.g., $\Delta_b$ as specified) after receiving the proposal before casting its vote, to allow slower nodes to catch up. In the worst case, the fastest honest node receives the block at time $t_0$ and the slowest at $t_0 + \Delta$ (with $\Delta \le \Delta_d$ in the parameters of our protocol). By the time the slowest honest node begins voting (after waiting $\Delta_b$), the fastest honest node has already broadcast its vote. All honest votes thus propagate and are received by every honest node within an additional $\Delta$ (bounded by $\Delta_v$). By the end of the allotted waiting time (on the order of $\Delta_d+\Delta_b+\Delta_v$ for the protocol parameters), every honest node has collected votes from all other honest nodes. A quorum of >$k/2$ consistent votes for the proposed block is reached, and all honest nodes will commit that block. No conflicting block can gather a quorum because either (a) there was a single proposal that all voted on, or (b) if a Byzantine leader tried to send divergent proposals (equivocation), honest nodes would detect the mismatch in block digests and immediately truncate the waiting period to vote against the faulty leader's block. In an equivocation scenario, honest votes will not favor two different blocks: at worst, honest nodes reject the leader's proposals altogether, resulting in no block for that slot rather than a fork. Additionally, if a malicious node attempts to cast two different votes, this double-voting is identifiable (each vote is signed by the node); honest nodes discard conflicting or invalid votes, so only one vote from each honest node counts. With an honest majority, no conflicting block can accumulate enough honest votes to be considered decided. Therefore, it is impossible for two different blocks to both reach commit threshold in the same shard, preserving safety.

    \emph{\textbf{Liveness}}: The automatic leader rotation ensures that even if the current slot's leader is Byzantine (e.g., fails to send a proposal or sends an invalid one), the protocol will not deadlock. If a leader equivocates or sends malformed data, honest nodes detect it (by invalid signatures or inconsistent digests) and treat that block as failed; the slot will pass with no commit. Thanks to frequent rotation, a new leader will be selected in the next slot. Given the honest majority, with overwhelming probability an honest leader will be chosen after at most a few rotations (indeed, the probability of continuously picking malicious leaders for many consecutive slots is negligible). When an honest leader eventually takes charge, they will propose a valid block and the above safety argument shows all honest nodes will commit it. Even in the case of a silence attack (a malicious leader simply does nothing), all honest nodes will detect the absence of a valid proposal by the end of the slot duration and move on to the next leader. Thus, a faulty leader may delay progress for at most one slot, but cannot prevent an honest leader in a subsequent slot from committing a block. As long as at least one out of any consecutive set of leaders is honest (which is guaranteed given $f<k/2$), the shard will continue to append new blocks. This guarantees liveness: every valid transaction will eventually be processed when it is proposed under an honest leader (or a malicious leader's turn passes without effect, and an honest leader later includes the transaction).

    In summary, the intra-shard consensus protocol maintains one consistent chain per shard (safety) and will always produce new blocks (liveness) under the stated assumptions. Byzantine behavior such as equivocation, duplicate voting, or silence are tolerated by design: honest nodes either circumvent the misbehavior within the same slot (e.g., by accelerating the voting upon detecting equivocation) or move to the next slot's leader, ensuring the protocol never violates safety and eventually commits all pending transactions.
\end{proof}

\subsection{Analysis of Cross-Shard Transactions}
\label{subsec:cross_analysis}

Finally, we formalize the security guarantees of our cross-shard transaction processing mechanism and discuss its overhead. Cross-shard transactions in SP-Chain are designed to achieve atomicity (the transaction executes in all involved shards or not at all, eventually) and integrity (no tampering or inconsistency in transaction data across shards), even under Byzantine participants. We assume the cryptographic primitives are secure (hash functions are collision-resistant and signatures are unforgeable) and each shard has an honest majority as before.

\begin{definition}[Atomicity of Cross-Shard Transactions]
    A cross-shard transaction protocol provides eventual atomicity if for every transaction involving a source shard (origin of funds/assets) and a destination shard, it guarantees that either both of the following hold in the long run: the source shard's state reflects the withdrawal (debit) and the destination shard's state reflects the corresponding deposit (credit), or neither shard's state reflects a completed transaction (no debit nor credit takes permanent effect). In simpler terms, funds or assets are neither lost nor duplicated: a cross-shard transfer will eventually complete in full, or be as if it never occurred.
\end{definition}

\begin{definition}[Integrity of Cross-Shard Transactions]
    Integrity means that the content of a cross-shard transaction cannot be altered or forged during its transfer between shards. The destination shard will only accept the deposit if it exactly matches the transaction that was confirmed in the source shard. Any attempt by an adversary to modify the transaction amount, recipient, or any associated data, or to introduce a fake cross-shard transaction, must fail verification and be rejected. This property ensures the consistency of transaction semantics across shards.
\end{definition}

\begin{thm}[Security of Cross-Shard Transaction Processing]
    In SP-Chain's cross-shard protocol, assuming honest majorities and secure cryptography, all cross-shard transactions satisfy atomicity and integrity. Specifically: (i) Atomicity: For any cross-shard transaction that debits an account in a source shard, the corresponding credit in the destination shard will eventually be committed (even if adversaries cause delays), so that the transaction's net effect is all-or-nothing. (ii) Integrity: No adversary can tamper with the transaction data or execute a cross-shard transfer illegitimately; the deposit will only be accepted if it is authorized by a confirmed withdrawal in the source shard (as proven by cryptographic proofs).
\end{thm}

\begin{proof}
    SP-Chain's cross-shard execution follows a two-phase approach (inspired by atomic commit protocols and prior shard designs \cite{wang2019monoxide, Harmony}): the transaction's withdrawal phase occurs in the source shard, then a deposit phase in the destination shard. Once the source shard commits the withdrawal (deducting the sender's funds and logging the intent to transfer), a cryptographic proof of this event is relayed to the destination shard. This proof includes the Merkle inclusion proof of the transaction in the source shard's block and the digital signatures attesting that block's validity. The security properties follow from this design:

    \emph{\textbf{Atomicity}}: If the withdrawal sub-transaction in the source shard fails to commit (e.g., the source leader never includes it, or the source consensus does not finalize it), then naturally no funds are deducted and the cross-shard transfer has no effect (the transaction can be retried or aborted without inconsistency). If the withdrawal does commit in the source shard, the protocol guarantees that the deposit will eventually commit in the destination shard. Even a malicious leader in the destination shard cannot permanently stall the deposit: because leaders rotate, an honest leader in that destination shard will eventually be elected and will process the pending deposit. At worst, a Byzantine leader might delay including the deposit, but they cannot cause the source's withdrawal to revert -- the funds are already moved out of the source account and are effectively held awaiting deposit. Clients are aware of the status (they receive confirmation from the source shard) and can inform the destination shard or retry the transfer message if needed. Thus, with an eventually honest leader and persistent retransmission, the deposit will be executed. The net result is that after a finite delay, the destination account is credited. The transaction is never half-completed: there is no scenario where the source deducted funds but the destination never receives them -- the worst-case outcome is a delay until an honest node finalizes the deposit. This meets the condition of eventual atomicity. (Likewise, it is impossible for a deposit to occur without its corresponding withdrawal: see integrity below.)

    \emph{\textbf{Integrity}}: The integrity is ensured by cross-shard verification using cryptographic proofs. When the destination shard receives the deposit request, it requires a proof that the corresponding withdrawal was confirmed in the source shard. This proof is the Merkle branch of the transaction in the source shard's block, along with that block's Merkle root and the source shard's block header (which contains signatures from the source shard's committee). Every node in the destination shard maintains an updated shard member table listing the public keys of all other shard committees (this table is refreshed each epoch during reconfiguration). Using this, the destination shard's nodes can authenticate the source block's signatures and thus trust the Merkle root. If a malicious actor alters the transaction (e.g., changing the amount or recipient in the deposit message), the Merkle proof will not match the altered data, and honest nodes in the destination shard will reject it. Similarly, an adversary cannot forge a valid proof for a transaction that never occurred in the source shard: without control of the majority of the source committee (which we assume they lack), they cannot produce a block header with a valid collective signature for a fabricated transaction. Any mismatch in signatures or Merkle roots is detected and the deposit is refused. Therefore, the only way a deposit is accepted is if it exactly corresponds to a legitimately confirmed withdrawal in the source shard. This guarantees that the state updates in the two shards are consistent and no cross-shard transaction can be modified or injected by Byzantine participants.
\end{proof}

}

\noindent\textbf{Communication Overhead Analysis.} 
Suppose a block contains a total of $N$ cross-shard transactions, where the number of cross-shard transactions sent to shard $j$ is $N_j$.  
In our design, since we organize the Merkle tree according to shards, we only need to send the root of each Merkle subtree to verify cross-shard transactions. 
Therefore, for each cross-shard transaction, the additional communication overhead is $O(m/N_j)$.  Correspondingly, in \cite{wang2019monoxide}, the extra communication overhead of each cross-shard transaction is $O(log_2(N))$.  
There might be thousands of transactions in a block, and most of them are cross-shard \cite{wang2019monoxide, zamani2018rapidchain,kokoris2018omniledger}.
However, there usually are only dozens of shards at most.
Therefore, our cross-shard processing scheme has a lower overhead under general cases.


%% file: evaluation.tex
\begin{figure*}[t]
\centering
\subfloat[Latency of broadcasting a digest.]{\includegraphics[scale=0.27]{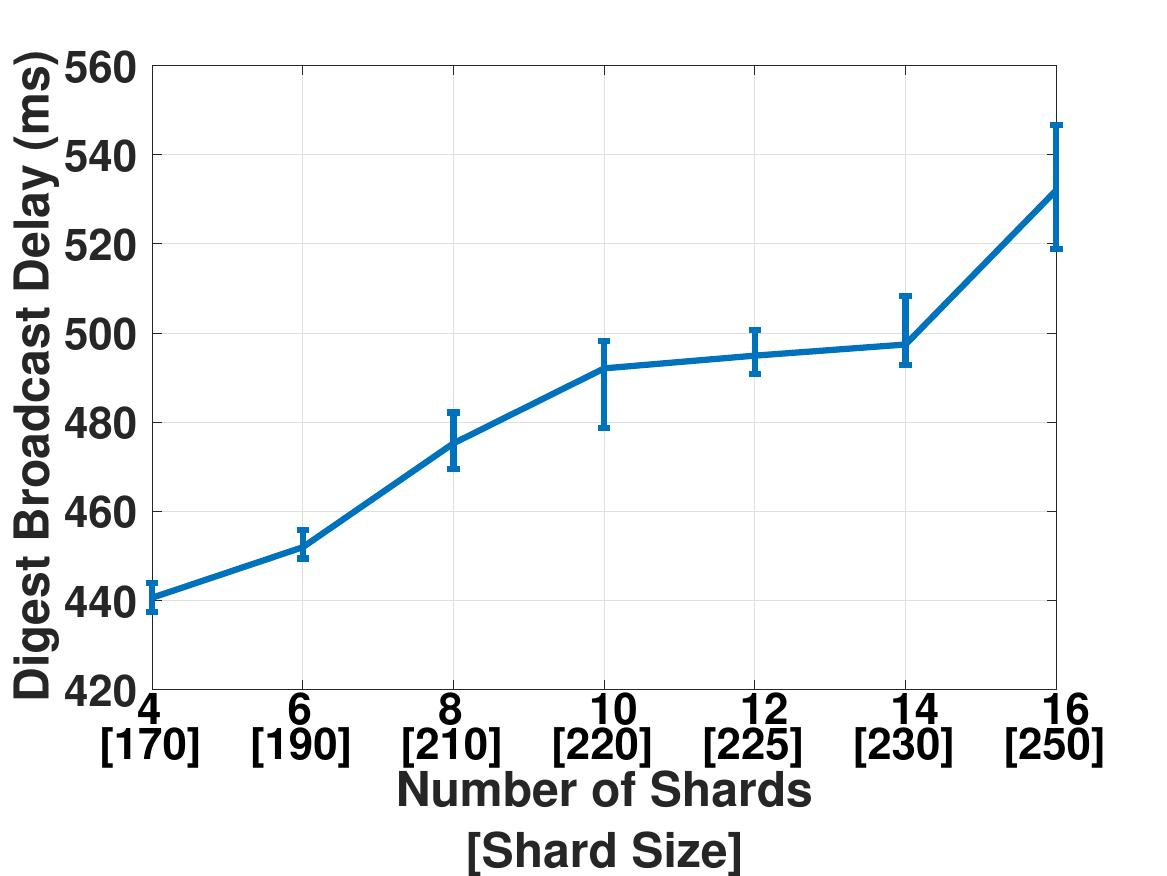}

}
\hfill
\subfloat[Latency of broadcasting a block (w/ digest).]{\includegraphics[scale=0.27]{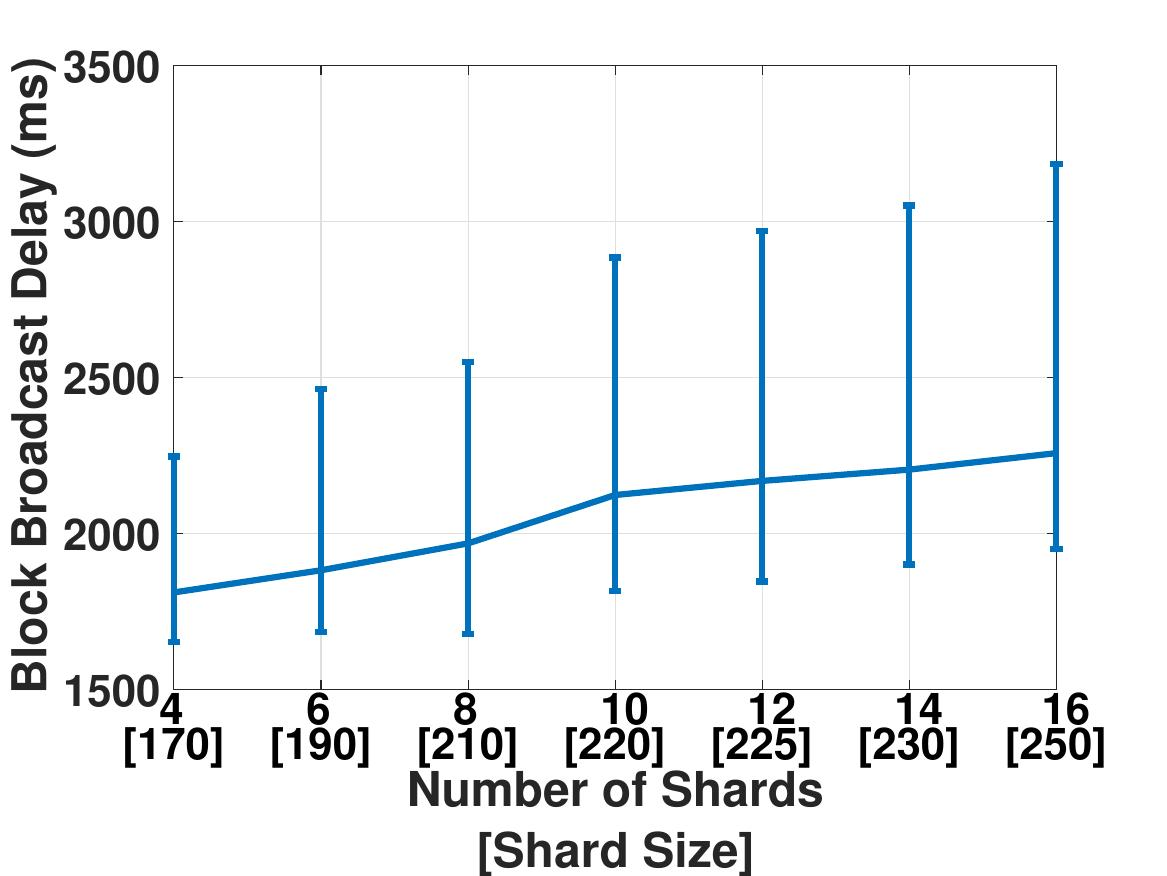}

}
\hfill
\subfloat[Latency of broadcasting a vote.]{\includegraphics[scale=0.27]{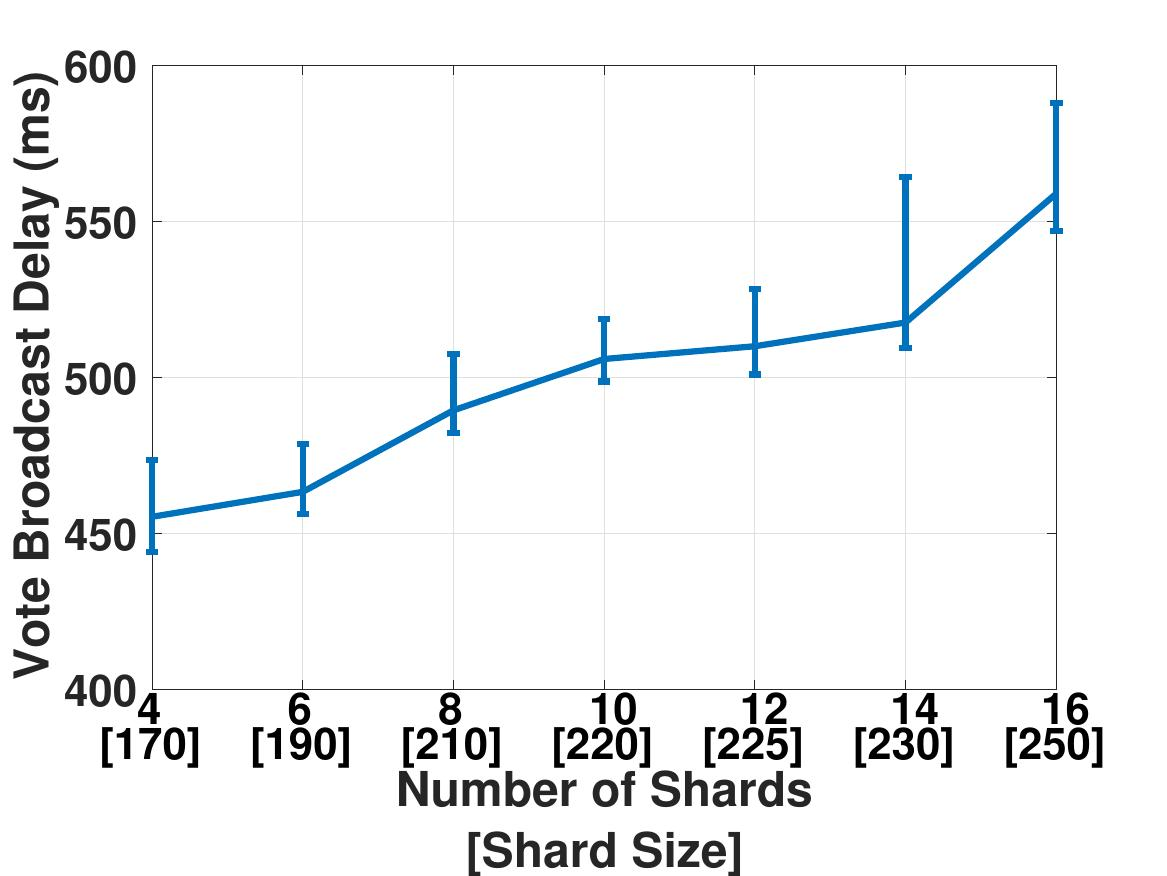}

}
\vspace{-4.5pt}
\caption{Latency of broadcasting a digest, a block (with digest), and a vote.}
\label{fig:delta}
\end{figure*}

\section{Implementation and Evaluation}
\label{sec:eva}

\subsection{Experimental Setup}
\label{sec:ExpSetup}

We implement SP-Chain based on Harmony \cite{Harmony} in Go language with 5,000+ lines of code.
We implement BLS aggregated signature \cite{boneh2001short} in the prototype to reduce the signature size for better performance.
For the baselines, we choose several notable existing blockchain sharding systems for performance comparison, including Monoxide \cite{wang2019monoxide}, OmniLedger \cite{kokoris2018omniledger}, and RapidChain \cite{zamani2018rapidchain}.
Actually, the main protocols in SP-Chain can be easily applied to most existing sharding systems to improve system performance. 


The choice of the experimental environment is similar to that in previous work \cite{zamani2018rapidchain}.
Specifically, we deploy SP-Chain on Amazon EC2 with up to 32 machines, each running up to 125 SP-Chain nodes. 
By default, the total network size scales up to 4,000 nodes.
{\color{red}
The machine is selected as c5.24xlarge, each with a 96-core Intel Xeon Platinum 8275L CPU, 192 GB memory, and a 25-Gbps communication link. 
It is worth noting that, since each node only maintains the state of its own shard and some metadata for cross-shard communication, the memory footprint per node remained modest.
To simulate geographically-distributed nodes, by default, we consider a latency of 100 ms for every message and a bandwidth of 20 Mbps for each node.
}
In each shard, we assume that when the adversaries observe the leader, they have the ability to attack it in the next slot.
Moreover, in each shard, the total malicious nodes are less than half of all nodes in a shard.
{\color{red}
We set each transaction size to 512 bytes, and each block contains up to 4,096 transactions, resulting in a block size of 2MB, similar to existing works \cite{zamani2018rapidchain}. 
The transactions are based on historical data of Ethereum \cite{zheng2020xblock}, in which the proportion of cross-shard transactions increases with the number of shards (e.g., when the number of shards is 16, the cross-shard transaction ratio is $15/16 = 93.75\%$).
}

\vspace{-12pt}
\begin{table}[tbh]
\caption{Choice of \# of nodes per shard and corresponding failure probability.}

\centering{}%
\begin{tabular}{c|c|c|c|c|c|c|c}
\hline

\hline
\# of shards & 4 & 6 & 8 & 10 & 12 & 14 & 16\tabularnewline
\hline 
\hline 
\# of nodes per shard & 170 & 190 & 210 & 220 & 225 & 230 & 250\tabularnewline
\hline 
Failure probability & \multirow{2}{*}{4.6} & \multirow{2}{*}{8} & \multirow{2}{*}{5} & \multirow{2}{*}{5} & \multirow{2}{*}{8} & \multirow{2}{*}{6} & \multirow{2}{*}{2}\tabularnewline
($\cdot10^{-7}$) &  &  &  &  &  &  & \tabularnewline
\hline

\hline
\end{tabular}
\label{table:ShardSize}
\end{table}

{\color{red}
\begin{figure}[t]
\centering
\includegraphics[scale=0.27]{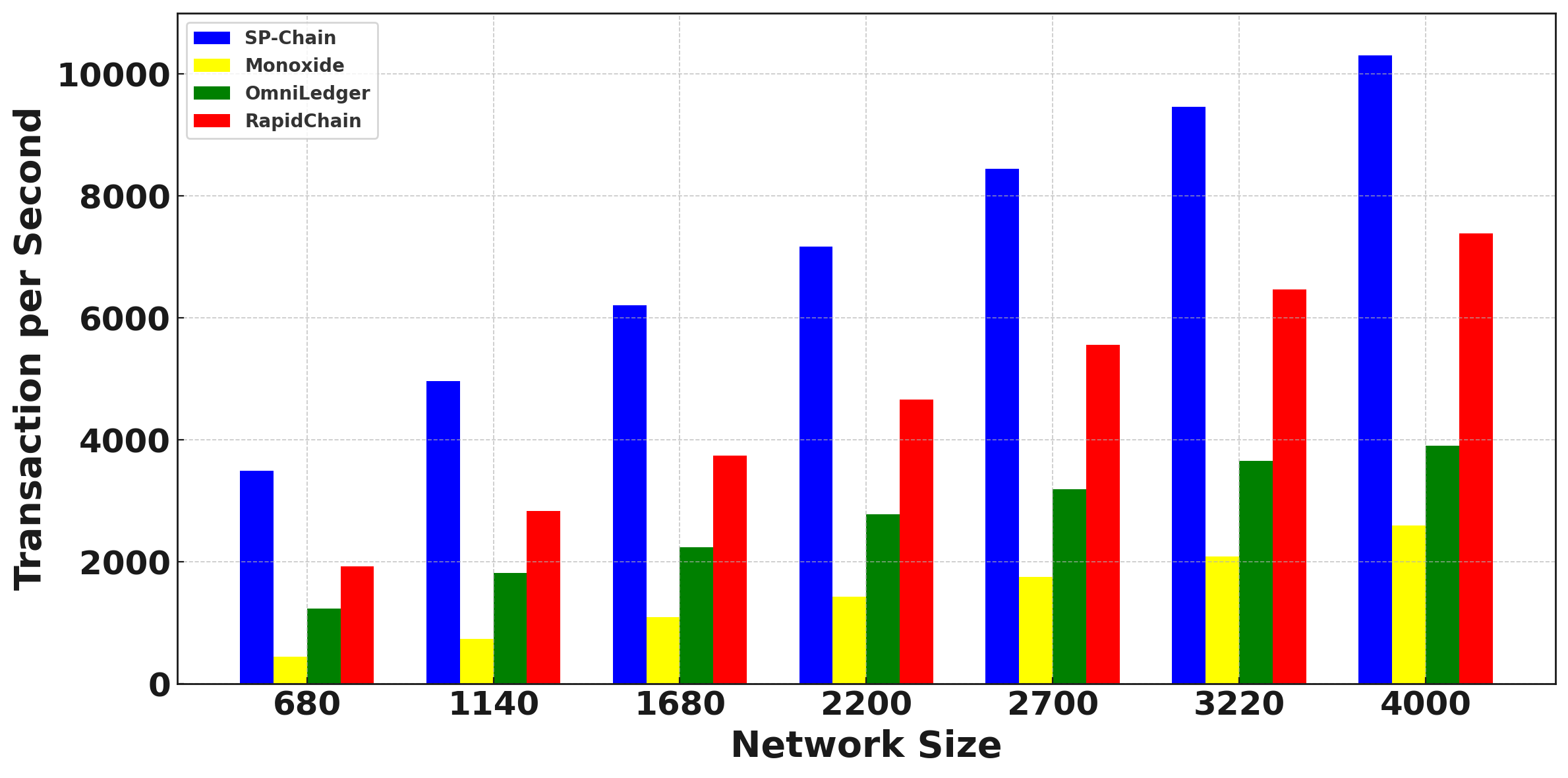}
\vspace{-12pt}
\caption{Throughput comparison results.}
\label{fig:TPSCompare}
\end{figure}

\begin{figure}[t]
\centering
\includegraphics[scale=0.27]{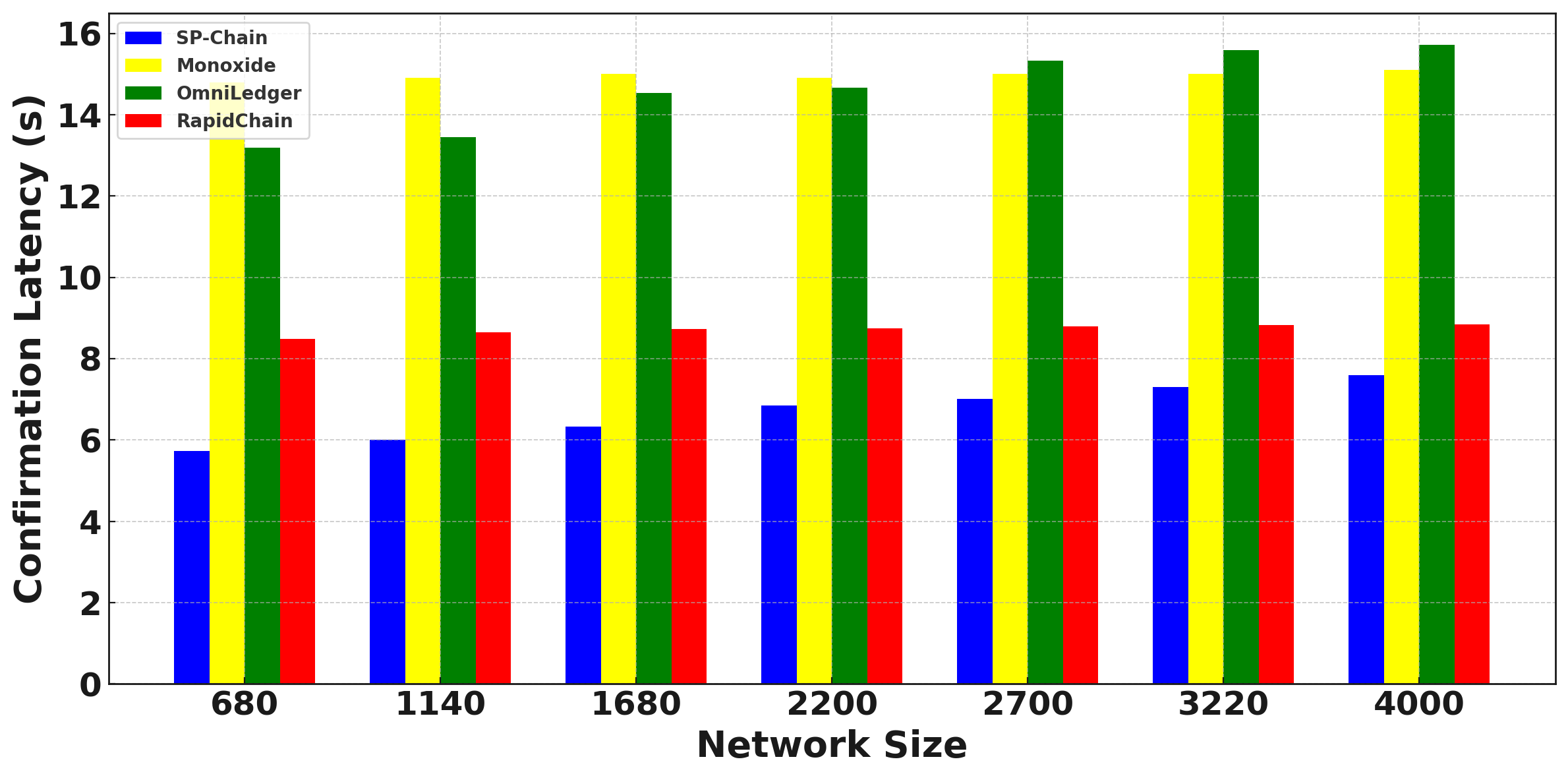}

\vspace{-12pt}
\caption{Transaction confirmation latency comparison results.}
\label{fig:Delay}
\end{figure}
}

\begin{figure}[t]
\centering
\includegraphics[scale=0.27]{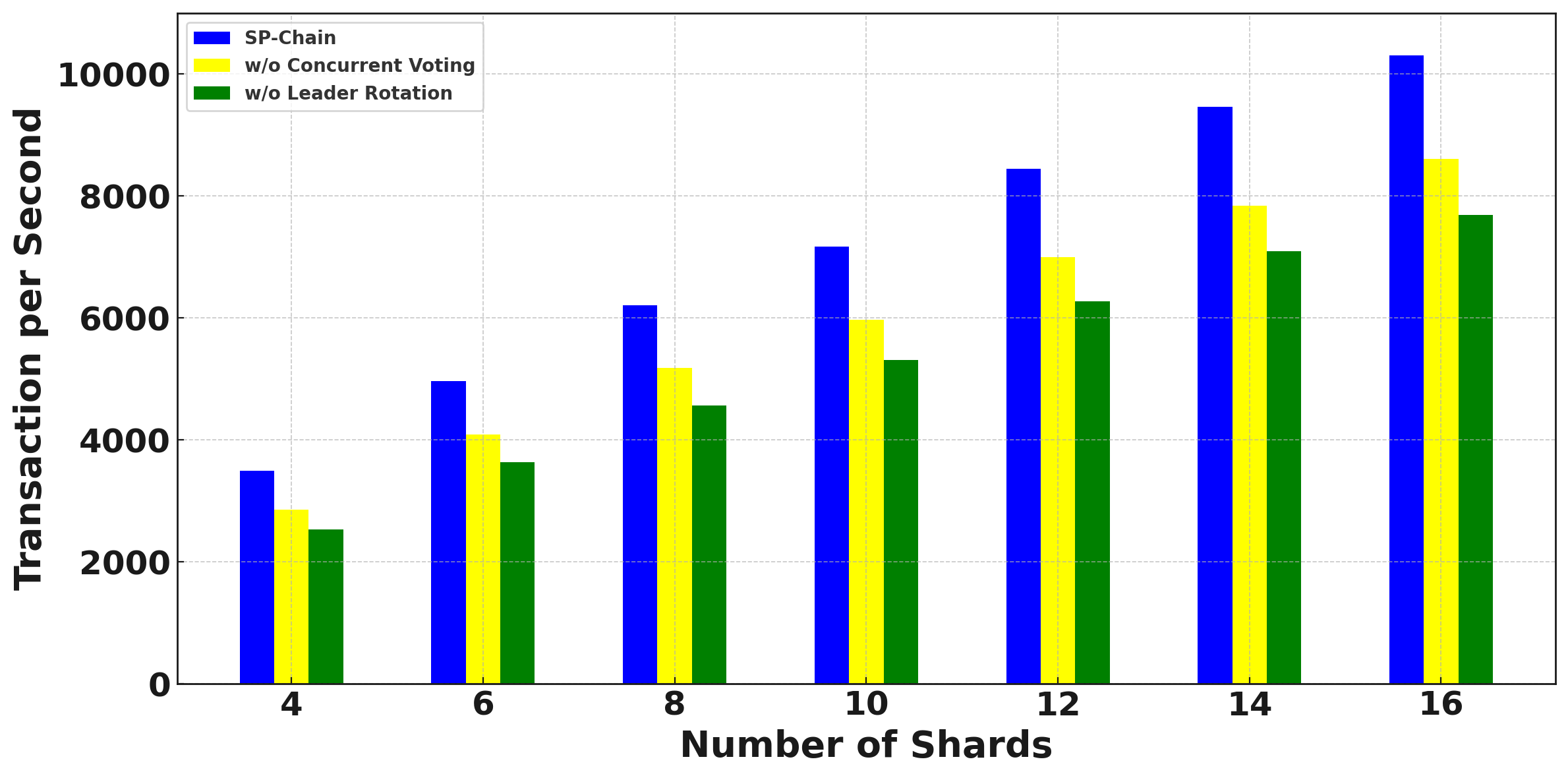}
\vspace{-12pt}
\caption{Throughput breakdown.}
\label{fig:TPSBreakdown}
\end{figure}

\noindent\textbf{Choice of Shard Size.}
We first determine the number of nodes per shard in SP-Chain based on Equation \ref{equ:failure}.
The rule for selecting the shard size is: the failure probability of each shard should be less than $2^{-20}\approx9\cdot10^{-7}$ (time-to-failure of more than 4580 years for one-day epoch) \cite{dang2019towards, zamani2018rapidchain}.
Table \ref{table:ShardSize} shows the choice of shard size under different shard numbers and the corresponding failure probability.  
Results show that our choice of shard size makes the probability of failure less than $9\cdot10^{-7}$ at any scale, ensuring the security of the system.
The following experiments will be conducted based on the shard size determined by Table \ref{table:ShardSize}.

\begin{table}[tbh]
\caption{Choice of $\triangle_{d}$, $\triangle_{b}$ and $\triangle_{v}$.}

\centering{}%
\begin{tabular}{c|c|c|c|c|c|c|c}
\hline

\hline
\# of shards & 4 & 6 & 8 & 10 & 12 & 14 & 16\tabularnewline
\hline 
\# of nodes & \multirow{2}{*}{170} & \multirow{2}{*}{190} & \multirow{2}{*}{210} & \multirow{2}{*}{220} & \multirow{2}{*}{225} & \multirow{2}{*}{230} & \multirow{2}{*}{250}\tabularnewline
per shard &  &  &  &  &  &  & \tabularnewline
\hline 
\hline 
$\triangle_{d}$ (ms) & 445 & 446 & 476 & 497 & 512 & 538 & 564\tabularnewline
\hline 
$\triangle_{b}$ (ms) & 2496 & 2737 & 2825 & 3236 & 3286 & 3397 & 3518\tabularnewline
\hline 
$\triangle_{v}$ (ms) & 541 & 542 & 581 & 616 & 626 & 651 & 683\tabularnewline
\hline

\hline
\end{tabular}
\label{table:delta}
\end{table}

\noindent\textbf{Choice of $\triangle$.}
The broadcast upper bound latency of digest, block and vote ($\triangle_{d}$, $\triangle_{b}$ and $\triangle_{v}$) are the 3 most important parameters in our system.  
To determine them, we evaluate the actual time spent in broadcasting digests, blocks, and votes in the system, and show the experimental results in Figure \ref{fig:delta}.  
Specifically, the bottom/top of each bar is the minimum/maximum delay in the broadcast, and the horizontal line connected to each bar is the average delay.  
Based on the measured actual broadcast delay, we set the broadcast delay upper bound as $delay_{max} + \sigma$.  
Where $delay_{max}$ is the actual maximum broadcast delay, $\sigma$ is the standard deviation obtained by fitting the broadcast delay to a normal distribution.  
Table \ref{table:delta} shows the calculated delay upper bounds $\triangle_{d}$, $\triangle_{b}$ and $\triangle_{v}$ under different shard sizes.  
By default, we will conduct experiments based on these values.





{\color{red}
\subsection{System Throughput}

We evaluate the system throughput and scalability of SP-Chain under varying network sizes from 680 up to 4,000 nodes, in direct comparison with Monoxide, OmniLedger, and RapidChain. Figure \ref{fig:TPSCompare} presents the throughput results across these scales. SP-Chain consistently achieves the highest throughput at every network size. 
For example, on a 680-node network SP-Chain sustains approximately 3,500 TPS, which already significantly exceeds the throughput of RapidChain (1,900 TPS), OmniLedger (1,200 TPS), and Monoxide (400 TPS). As the network size grows, this gap widens. 
Under the largest tested scale of 4,000 nodes, SP-Chain processes 10,305 TPS, whereas RapidChain and OmniLedger reach only about 7,400 TPS and 3,900 TPS respectively, and Monoxide remains 2,600 TPS. 
In other words, at 4,000 nodes SP-Chain achieves roughly $1.4\times$ the throughput of RapidChain, about $2.6\times$ that of OmniLedger, and around $4\times$ that of Monoxide. 
Furthermore, SP-Chain's throughput increases almost linearly with the expansion of network size, demonstrating superior scalability. 


\subsection{Transaction Latency}

We now compare the average transaction confirmation latency of SP-Chain against baselines as the network size increases. Figure \ref{fig:Delay} shows the confirmation delay results for network sizes from 680 to 4,000 nodes. 
Across all scales, SP-Chain achieves substantially lower latency than the other protocols. 
At the 680-node scale, a transaction in SP-Chain is confirmed in roughly 6s, whereas RapidChain requires about 8s, OmniLedger about 13s, and Monoxide around 15s on average for confirmation. 
As the network grows, the confirmation latency of every protocol rises due to larger shard sizes and increased cross-shard coordination. 
SP-Chain's latency, however, grows very modestly and remains the lowest among the systems. 
In a 4,000-node network, SP-Chain confirms transactions in about 7.6s, which is roughly half of that in OmniLedger and Monoxide, and about 15\% lower than in RapidChain. 
These results demonstrate that SP-Chain sustains low confirmation latency even as the system scales, outperforming prior protocols on this metric. 
Combining throughput and latency, our evaluation shows that SP-Chain does not trade off latency for throughput -- it delivers both higher throughput and faster confirmations than various baselines across the range of network sizes tested.

}

\subsection{System Decomposition}
In this section, we decompose SP-Chain and evaluate the impact of different system components on the throughput in detail.
We mainly analyze how much the system throughput is improved by concurrent voting and leader rotation.

\vspace{3pt}
\noindent\textbf{Concurrent Voting and Leader Rotation.}  
Figure \ref{fig:TPSBreakdown} shows the SP-Chain throughput results after removing the concurrent voting or leader rotation mechanism.  
The yellow/green bar indicates the case that concurrent voting/leader rotation is removed from SP-Chain.
The results show that the concurrent voting mechanism can increase throughput by up to 22\%.  
The leader rotation mechanism increases the throughput by up to 37\%. 
This is mainly because, with the leader rotation mechanism, the system eliminates the view change process.
{\color{red}In cases} where the leaders are attacked by adversaries, the system's efficiency is thus greatly increased.
Under the network scale of 4,000 nodes, the concurrent voting and leader rotation mechanism increases the system throughput by 20\% (10,305 vs 8,612) and 34\% (10,305 vs 7,692), respectively.  
In summary, the concurrent voting and leader rotation mechanism improves the throughput of the system significantly.

\vspace{-12pt}
\begin{table}[tbh]
\centering
\caption{Performance of SP-Chain at 4,000 and 10,000 nodes.}
\begin{tabular}{c| c| c| c| c}
\hline

\hline
\textbf{Shards} & \textbf{Nodes/Shard} & \textbf{Failure Prob.} & \textbf{Latency (s)} & \textbf{TPS} \\
\hline
\hline
16 & 250 & $2\times10^{-7}$ & 7.6 & 10,305 \\
\hline
40 & 250 & $5\times10^{-7}$ & 8.1 & 24,509 \\
\hline

\hline
\end{tabular}
\label{table:large_scale}
\end{table}

{\color{red}
\subsection{Scalability Beyond 4,000 Nodes}

To further evaluate SP-Chain's scalability beyond 4,000 nodes. 
We deploy the system in a larger network of 10,000 nodes, composed of 40 shards with 250 nodes per shard. 
The results of this large-scale experiment are summarized in Table \ref{table:large_scale}. SP-Chain continues to deliver high throughput and low latency at 10,000 nodes: the end-to-end transaction confirmation latency is 8.1 seconds, while throughput reaches 24,509 transactions per second. The system's failure probability remains extremely low, at approximately $5\times10^{-7}$ (per epoch), which is well below the $9\times10^{-7}$ benchmark used to ensure negligible failure risk. 
Comparing to the 4,000-node case, the 10,000-node deployment demonstrates near-linear scaling. Throughput increases by about $2.4\times$ (24,509 vs 10,305) with only a modest rise in confirmation delay (8.1s vs 7.6s). These results validate that SP-Chain can scale to large networks with minimal performance degradation, confirming that the system maintains its high performance even beyond 4,000 nodes.

\begin{figure}[t]
\centering
\includegraphics[scale=0.27]{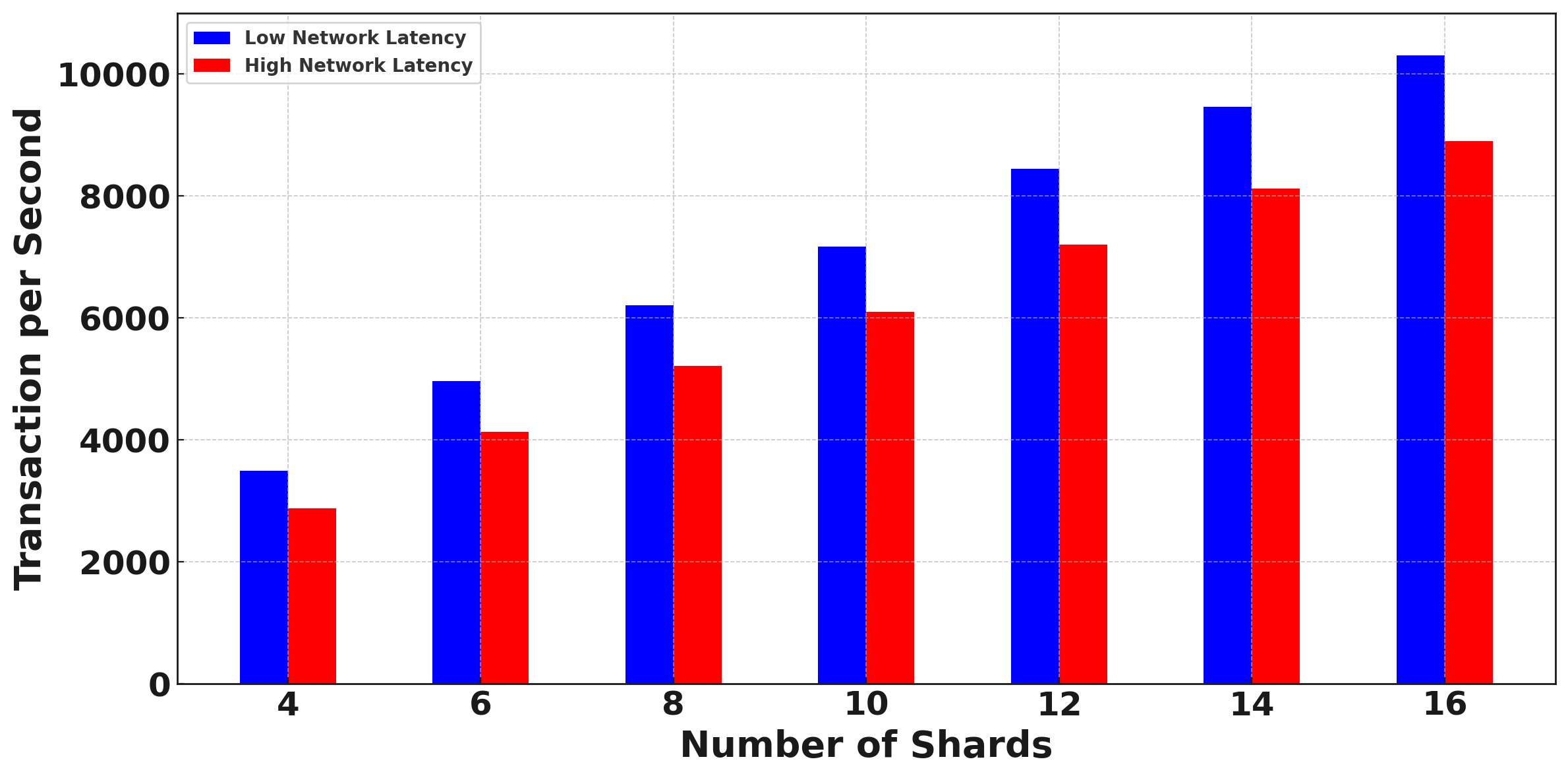}
\vspace{-12pt}
\caption{Throughput under different network latency.}
\label{fig:TPS_high_latency}
\end{figure}

\begin{figure}[t]
\centering
\includegraphics[scale=0.27]{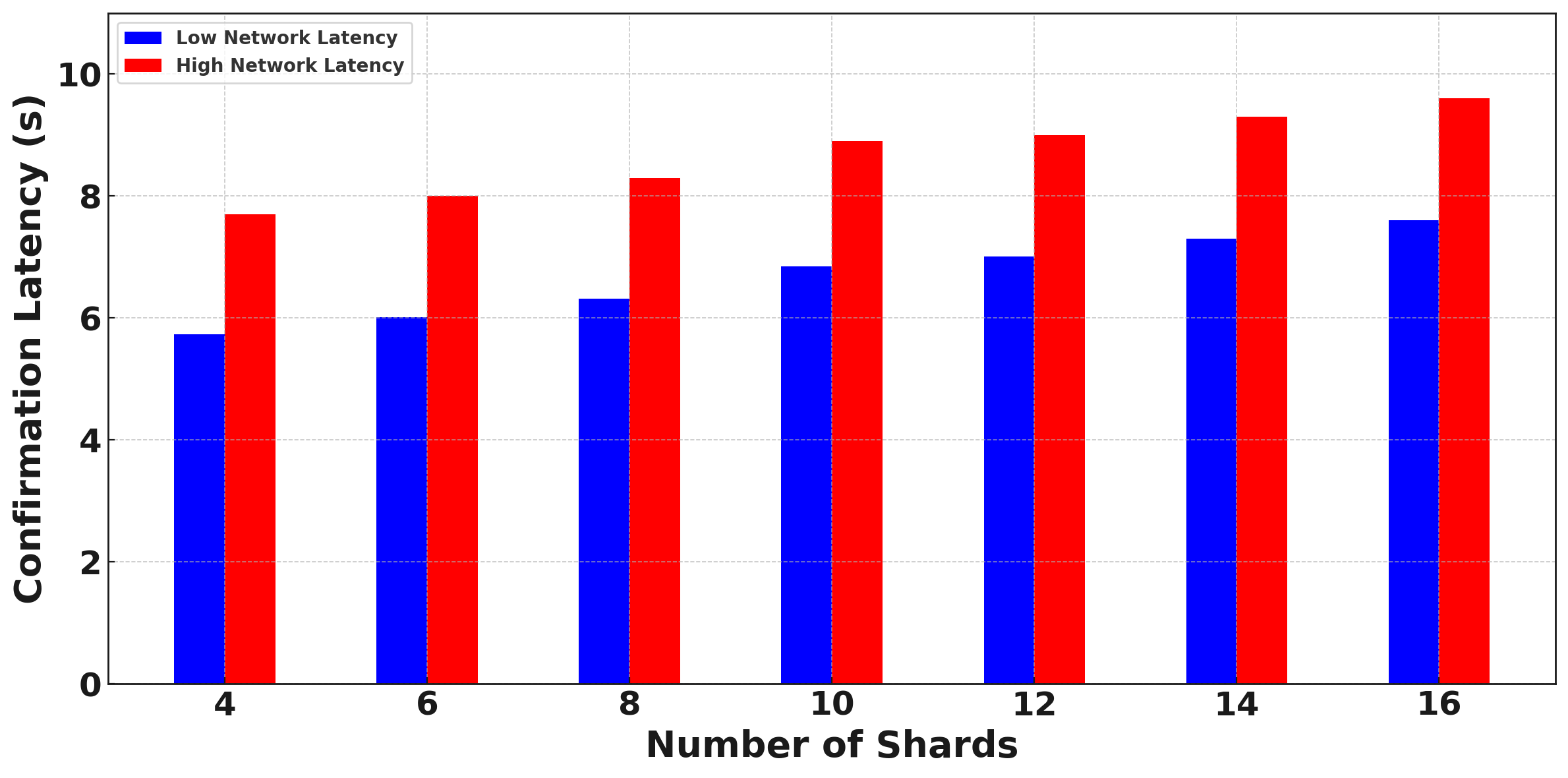}

\vspace{-12pt}
\caption{Transaction confirmation latency under different network latency.}
\label{fig:Delay_high_latency}
\end{figure}

\subsection{Impact of Network Latency}

To evaluate SP-Chain under adverse network conditions (e.g., network churn with fluctuating delays), we simulate a high-latency environment. Specifically, each inter-node message is delayed by a random interval between 100 ms and 1000 ms. We compare SP-Chain's performance in this high-latency scenario to the default low-latency case (100 ms) across various shard counts. 

Figure \ref{fig:TPS_high_latency} presents the throughput results, with blue bars for the baseline and red bars for the high-latency case. 
Under high network latency, the throughput is only modestly reduced compared to the low-latency case. For instance, at 16 shards the system achieves around 8,900 TPS with high delay, approximately 14\% lower than the 10,305 TPS under low-latency conditions. 
This minor degradation indicates that SP-Chain's pipelined consensus and concurrent block processing effectively tolerate substantial message latency jitter. The consensus mechanism continues to produce blocks at a high rate, highlighting robust performance even when network delays fluctuate by an order of magnitude.

The high-latency scenario's effect on transaction confirmation time is likewise well-contained. 
Figure \ref{fig:Delay_high_latency} shows the median transaction confirmation latency (time to finality) for SP-Chain under both low and high network latency conditions. 
As expected, introducing large message delays leads to higher confirmation times, but the impact is moderate. 
On average, the confirmation latency rises by only about 2 seconds when moving from the baseline to the high-latency scenario. 

}

%% file: Discussion.tex
{\color{red}
\section{Discussion and Future Directions}

\subsection{Potential Attack Vectors and Mitigations}

We now enumerate some attack vectors and concisely explains how SP-Chain neutralizes each threat.

\vspace{3pt}
\noindent
\textbf{Cross-Shard Replay Attacks. }
In SP-Chain, cross-shard transactions are protected by our proof-based verification mechanism that prevents replay across shards. Each cross-shard transaction must carry a cryptographic proof of its inclusion in the source shard's ledger (as described in Section \ref{sec:cross-shard}). This batched Merkle proof ties the transaction to a specific confirmed block and state, allowing the destination shard to verify its validity before acceptance. A transaction replayed without the correct proof (or with a proof already used) will be rejected by the receiving shard, ensuring that cross-shard replay attacks are effectively mitigated.

\vspace{3pt}
\noindent
\textbf{Randomness Manipulation via Block Withholding. }
SP-Chain's leader election protocol is designed to be unbiased and resistant to manipulation (Section \ref{sec:leader}). A new leader is chosen for every block using a random seed derived from the previous block's committed signatures. An adversarial leader cannot influence this process without aborting their own block proposal. If a leader attempts to withhold a block to sway the randomness, the protocol simply rotates to the next leader for the subsequent slot. With automatic per-block leader rotation and no gain from stalling, block withholding yields no advantage in biasing the random seed, thwarting this attack vector. 

\vspace{3pt}
\noindent
\textbf{Adaptive Adversaries and Leader Selection Biases. }
By design, SP-Chain prevents adaptive adversaries from exploiting leader selection. Leaders are rotated frequently and randomly, so no node remains leader beyond a single block. The next leader is determined just before each block using unpredictable distributed randomness, meaning attackers cannot know or influence who comes next (as detailed in Section \ref{sec:leader}). This ephemeral and random leadership schedule leaves adversaries with minimal time to adapt or target specific leaders, eliminating any consistent bias that could be exploited in the leader selection process. Even a well-resourced attacker cannot reliably preempt or co-opt the leader, which preserves the protocol's fairness and security against adaptive strategies. 

\vspace{3pt}
\noindent
\textbf{Censorship Attacks. }
SP-Chain mitigates transaction censorship through rapid leader turnover and verifiable cross-shard communication. Because leadership changes every block, a malicious leader cannot continuously suppress a transaction--any transaction omitted in one block can be included by the next leader once the adversary's turn ends. Moreover, the cross-shard transaction scheme makes any censorship attempt evident. If a leader tries to drop or alter a cross-shard transfer, the inconsistency will be detected when the destination shard recomputes the Merkle root and compares it against the signed block header. Honest nodes would notice the mismatch and refuse to acknowledge the tampered data, forcing eventual inclusion of the legitimate transaction. Thus, by design (see Section \ref{sec:cross-shard}), SP-Chain ensures that censorship is at most temporary and cannot persist undetected or uncorrected.

\subsection{Real-World Deployment and Limitations}

We now summarize the practical considerations for deploying SP-Chain, while highlighting how the protocol's core mechanisms address each challenge.

\vspace{3pt}
\noindent
\textbf{Integration with Existing Ecosystems. }
SP-Chain's design can be integrated with contemporary blockchain architectures. Our Harmony-based prototype demonstrates that SP-Chain's protocols are easily applicable to existing sharded networks. For example, Ethereum 2.0's multi-shard architecture could adopt SP-Chain's pipelined consensus and unbiased leader rotation to improve throughput and fairness. Similarly, SP-Chain's efficient cross-shard transaction processing would enable seamless asset transfers across shards in a multi-chain ecosystem like Ethereum. 

\vspace{3pt}
\noindent
\textbf{DeFi Use Cases and Regulatory Fairness. }
For DeFi applications, which demand high throughput and fairness, SP-Chain offers low-latency processing (as shown in Section \ref{sec:eva}) and an unbiased leader election protocol that helps mitigate front-running. Front-running--where insiders exploit transaction ordering--is illegal in traditional markets, yet it remains a challenge in DeFi; SP-Chain's fair leader rotation makes such exploits significantly harder on-chain. By preventing any single validator from consistently controlling transaction ordering, SP-Chain aligns with regulatory expectations for market fairness and bolsters trust in decentralized exchanges and financial platforms. 

\vspace{3pt}
\noindent
\textbf{Cross-Border Payments and Enterprise Blockchains. }
Beyond DeFi, SP-Chain is applicable to cross-border payment networks and enterprise blockchains. Global payment systems can leverage SP-Chain's high scalability for fast, low-cost international transactions; shards might correspond to different currencies or regions, with SP-Chain's cross-shard mechanism ensuring secure atomic transfers across them. Similarly, enterprise blockchain deployments can adopt SP-Chain in a permissioned setting by partitioning consortium participants into shards for parallel processing. The protocol's Byzantine fault tolerance and rotating leadership guarantee that no single organization can monopolize block production, preserving fairness and performance in an enterprise context. 

\vspace{3pt}
\noindent
\textbf{Energy Efficiency Trade-offs. }
SP-Chain is designed to be energy-efficient: instead of proof-of-work, it uses a Byzantine fault-tolerant consensus with finality, avoiding wasteful mining. By processing transactions in parallel across shards, nodes handle only a fraction of the total workload, reducing redundant computation and lowering the overall energy cost per transaction relative to a single-chain system. Coordinating multiple shards does introduce some communication overhead; however, this cost is offset by throughput gains, keeping energy per transaction comparable to (or lower than) an unsharded architecture while delivering much higher total throughput. 

\vspace{3pt}
\noindent
\textbf{Deployment Challenges and Limitations. }
Despite its benefits, SP-Chain faces several real-world deployment challenges. Integrating SP-Chain into an existing blockchain may require bridging infrastructure or consensus modifications; however, many issues (e.g., malicious leader handling and cross-shard security, addressed in Sections \ref{sec:leader} and \ref{sec:cross-shard}) are intrinsically solved by its design, easing deployment in adversarial settings. A remaining limitation is the reliance on timely network communication--if network latency exceeds expected bounds, consensus efficiency could degrade. SP-Chain also depends on a secure source of distributed randomness for leader selection, which must be managed in production. Finally, broader concerns like governance and regulatory compliance (for example, identity management in enterprise shards or legal frameworks for cross-border use) lie outside the protocol's scope and must be addressed at the ecosystem level.

\subsection{Future Directions and Research Opportunities}

One promising direction is to leverage machine learning to enhance the shard leader selection process. SP-Chain's current leader rotation is random and unbiased to ensure fairness and security, but incorporating data-driven or learning-based strategies could further boost efficiency. For example, a reinforcement learning model or reputation scoring system could dynamically select leaders based on past performance and reliability \cite{hussain2025reputation}. By favoring nodes with consistently high throughput and honest behavior, the system might reduce downtime or stalls caused by suboptimal leaders, thus improving overall sharding efficiency without compromising security. 

Another critical research avenue is integrating post-quantum cryptography into SP-Chain's cross-shard transaction mechanism. While SP-Chain currently relies on classical cryptographic assumptions (e.g., digital signatures and hash-based proofs) for cross-shard security, these could become vulnerable with the advent of quantum computing. To future-proof the system, one can replace or augment these primitives with post-quantum cryptosystems \cite{gharavi2024post}. For instance, using quantum-resistant digital signatures or key exchange protocols (such as lattice-based or multivariate polynomial schemes \cite{wang2021low}) would ensure that cross-shard transaction validation remains secure against quantum-capable adversaries. Adopting such cryptographic upgrades would not require fundamental changes to the SP-Chain protocol, thereby preserving its performance while extending its security longevity. 

Finally, we highlight the broader real-world impact of SP-Chain and the new research opportunities it enables. By achieving high throughput and robust security in a sharded blockchain, SP-Chain serves as a practical blueprint for scaling distributed ledgers in real-world scenarios (e.g., financial networks or large-scale IoT deployments). Its design principles and techniques can inspire further innovation beyond our specific implementation. For example, other researchers and practitioners could adapt SP-Chain's concurrent consensus or cross-shard proof mechanisms to improve interoperability between different blockchain platforms, or to build next-generation applications that demand both scalability and security. 
}